\documentclass[aps,preprint,showpacs,showkeys]{revtex4}

\usepackage{amsmath}
\usepackage{amssymb}
\usepackage{graphicx}
\usepackage{color}
\usepackage{verbatim}

\newcommand{\dpr}[0]{$\Delta^{\prime}$}

\begin{document}

{\raggedleft {\it Accepted for publication on Physics of Plasmas}}
\\

\title{Numerical comparison between a Gyrofluid and Gyrokinetic model investigating collisionless magnetic reconnection}

\author{O. Zacharias$^{1}$, L. Comisso$^{2,3}$, D. Grasso$^{2,3}$, R. Kleiber$^{1}$, M. Borchardt$^{1}$ and R. Hatzky$^{4}$ \\
{\small $^1$ IPP-Teilinstitut Greifswald, Wendelsteinstr. 1, D-17491 Greifswald, Germany   } \\
{ {\small $^2$ Dipartimento Energia, Politecnico di Torino, Corso Duca degli Abruzzi 24, 10129, Torino, Italy} \\  
{\small $^3$ Istituto dei Sistemi Complessi - CNR, Via dei Taurini 19, 00185, Roma, Italy} \\
{\small $^4$ Max-Planck-Institut f{\"u}r Plasmaphysik, Boltzmannstr.~2, 85748 Garching, Germany } \\}  }

\baselineskip 24 pt

\begin{abstract}

The first detailed comparison between gyrokinetic and gyrofluid simulations of collisionless magnetic 
reconnection has been carried out. Both the linear and nonlinear evolution of the collisionless tearing 
mode have been analyzed. In the linear regime, we have found a good agreement between the two approaches 
over the whole spectrum of linearly unstable wave numbers, both in the drift kinetic limit and for finite 
ion temperature. Nonlinearly, focusing on the small-$\Delta '$ regime, with $\Delta '$ indicating the 
standard tearing stability parameter, we have compared relevant observables such as the evolution and 
saturation of the island width, as well as the island oscillation frequency in the saturated phase.
The results are basically the same, with 
small discrepancies only in the value of the saturated island width for moderately high values
of $\Delta '$. Therefore, in the regimes investigated here, the gyrofluid approach can describe the 
collisionless reconnection process as well as the more complete gyrokinetic model.

\end{abstract}



\maketitle

\section{Introduction}

Reconnection of magnetic fields is recognized to play a key role in many events occurring in laboratory, 
space, and astrophysical plasmas. Classical examples of such events are sawtooth crashes in tokamaks, 
substorms in the Earth's Magnetosphere, and solar flares. Magnetic reconnection involves a topology 
change of a set of field lines, which leads to a new equilibrium configuration with lower magnetic energy. 
During this process magnetic energy is converted into kinetic and thermal energy of electrons and ions~\cite{YKJ_2010}. 
Although much of the progress in the understanding of magnetic reconnection has been possible thanks to the 
use of fluid-based models, the results achieved with these models require independent confirmation when kinetic 
effects are expected to be important. 

Recently, a new class of generalized fluid models, so called gyrofluid models, have been adopted to investigate 
magnetic reconnection in the presence of a large guide field 
~\cite{GCPP_2000,LH_2008,GTW_2010,TWG_2010,DMPC_2011,BS_12,CGTW_2012,CWG_2012,CGWB_2013}. 
These models combine 
the advantages of the fluid description, namely computational efficiency and intuitively appealing physical 
interpretation, while retaining important kinetic effects through gyro-orbit averaging~\cite{Waelbr_2011}. However, 
for problems in which strongly non-Maxwellian features characterize the
distribution function it would probably be necessary to keep many velocity-space moments to describe the detailed 
shape of the distribution function, in which case the gyrofluid approach may lose its advantages~\cite{DH_1993}. 
Therefore, detailed comparisons between gyrokinetic and gyrofluid simulations are necessary to confirm the 
validity of the continuum gyrofluid descriptions and to improve them when such descriptions are no longer applicable.

An early investigation of collisionless tearing modes by means of gyrokinetic particle-in-cell (PIC) simulations 
was made in Ref.~\cite{Sydora_2001}. In particular, this work focused on the growth and nonlinear evolution of
small-scale magnetic islands having a characteristic width of the order of the electron skin depth and smaller 
than the ion Larmor radius. In Ref.~\cite{WCP_2005} the evolution of collisionless and semicollisional tearing 
mode instabilities was studied using a gyrokinetic $\delta f$ PIC code with gyrokinetic ions and drift-kinetic 
electrons. After a benchmark of the linear simulation results with eigenmode analysis for the case of fixed ions, 
 the nonlinear evolution of the magnetic island width was calculated. More recently, in Ref.~\cite{ZKH_2012} numerical 
results of saturated island widths resulting from gyrokinetic $\delta f$ PIC simulations were compared to 
analytical calculations~\cite{DL_1977_PRL} in a more extended parameter space. In this work electron diamagnetic 
effects were also considered, and it was found that they have stabilizing effects in agreement with the asymptotic 
theory of Ref.~\cite{P_1991}. Simulations of the collisionless tearing mode
with gyrokinetic electrons and fully kinetic ions were performed in Ref. 
~\cite{Wang_2011} and compared with the asymptotic 
matching theory of Ref.~\cite{DL_1977_PRL}, and with a gyrokinetic 
eigenmode theory in a small but finite Larmor radius limit.  
{Very recently, collisionless reconnection in the large guide field regime 
has also been investigated by comparing fully kinetic PIC simulations and gyrokinetic results, 
showing that the gyrokinetic framework is capable of making accurate predictions 
well outside its formal regime of applicability~\cite{TenBarge_2013}. } 
It was also shown that many physical quantities resulting from the nonlinear reconnection process scale linearly with the guide field.

The first comparison between gyrokinetic and fluid simulations was carried out in~
Ref.~\cite[]{Rogers_2007, Rogers_2007_2}, where both 
the linear and nonlinear regimes of collisionless magnetic reconnection were investigated, finding a reasonably good 
agreement between the two approaches for low-$\beta$ plasmas and small ion to electron temperature ratio. 
For $\beta \sim 1$ and ion temperature greater than the electron 
temperature, an increase in discrepancy between gyrokinetic simulations and fluid theory was found
in Ref.~\cite{Numata_2011}, where, however, it was shown that the adoption of a reduced 
ion-to-electron mass ratio plays a significant role in causing these discrepancies. This latter work focused on the 
linear regime, but also considered the collisionality dependence of the tearing mode growth. The importance of 
adopting a realistic mass ratio was emphasized in Ref.~\cite{Pueschel_2011}, where extensive linear studies were 
presented, and nonlinear results were performed to investigate reconnection in the cases of decaying and driven turbulence.\\
As in most of the works mentioned above, in this paper we focus on rarefied high-temperature plasmas in which the 
collisional mean free path is large enough that collisions are negligible. Additionally we consider magnetic 
reconnection phenomena that take place in a two-dimensional plane perpendicular to a strong and constant magnetic guide 
field. Differently from previous studies, here the comparison is between the results of gyrokinetic and gyrofluid 
simulations. For this purpose, we adopt the gyrokinetic $\delta f$ PIC code EUTERPE with gyrokinetic ions and 
drift-kinetic electrons~\cite{ZKH_2012}. 
Recently, a linear version of this code (GYGLES) has been employed to simulate the ideal-MHD 
internal kink mode and the collisionless $m=1$ tearing mode in a tokamak~\cite{MZ_2012}. 
We also adopt the gyrofluid code that has been 
employed in Refs.~\cite{CGTW_2012,CWG_2012,CGWB_2013} to investigate ion gyro-orbit averaging effects on collisionless 
magnetic reconnection. After a linear benchmark of these codes with a numerical eigenmode and eigenvalue analysis, the 
results of the two models in the linear regime are compared over the whole spectrum of linearly unstable wave numbers, 
both in the drift kinetic limit and for finite ion temperature. Nonlinearly, focusing on the small $\Delta '$ regime 
(with $\Delta '$ indicating the standard tearing stability parameter), we compare relevant observables as the evolution 
and saturation of the island half-width, and the {island oscillation} frequency at saturation.

This paper is organized as follows: In Sec.~\ref{sec2} the adopted gyrokinetic and gyrofluid models are described, as 
well as the initial equilibrium configuration. In Sec.~\ref{sec3} we focus on linear simulation results, while the 
nonlinear regime is studied in Sec.~\ref{sec4}. Finally, in Sec.~\ref{sec5} we summarize our results and discuss their 
implications.

\section{The models}  \label{sec2}

Within the framework of low-$\beta$ plasmas, $\beta \ll 1$, the dominant field fluctuations
 are the electrostatic potential $\hat{\Phi}$ and the parallel vector potential $\hat{A}_{\|}$. 
Both models which are investigated here adopt the following normalization scheme with respect to Alfv{\'e}n units
\begin{eqnarray}
t = \frac{v_{\rm A}}{L}\,\hat{t},\quad x = \frac{\hat{x}}{L}, 
\quad d_s = \frac{\hat{d}_s}{L} , \quad \rho_{{\rm S},e} = \frac{\hat{\rho}_{{\rm S},e}}{L}, \\
\quad n_s = \frac{L \, \hat{n}_s}{\hat{d}_i\, n_0}, \quad u_s = \frac{L \, \hat{u}_s}{\hat{d}_i\, v_{\rm A}}, 
\quad A=\frac{\hat{A}_{\|}}{B_{0,z} \, L}, \quad \Phi=\frac{\hat{\Phi}}{B_{0,z} \, L\, v_{\rm A}} 
\label{normalization}
\end{eqnarray}
where the carets denote the dimensional quantities,
$u_s$ is the out-of-plane guiding center velocity field, $n_s$ is the guiding center density perturbation,
and a constant background density $n_{\textrm{0}}$ is assumed to be equal for each species $s$.
$L$ indicates a characteristic 
magnetic equilibrium length scale, { while $v_{\rm A}=B_{0,z}/\sqrt{\mu_0\,n_0\,m_i}$ is the Alfv{\'e}n speed based 
on the magnetic field strength $B_{0,z}$ of the guiding field.} 
$\hat{d}_s=c/\omega_{{\rm p},s}$ is the skin depth of singly charged ions ($s=1$) or electrons ($s=2$)
and $\rho_{{\rm S},e}=\sqrt{m_i\,k_{\rm B}\,T_{0,e}}/\left(e\,B_{0,z}\right)$ is the sound Larmor radius. 
The ratio of ion temperature 
$T_{0,i}$ to the reference temperature of the electrons $T_{0,e}$ is indicated by $\tau$,
while $\mu$ refers to the ratio of the ions mass $m_i$ to electron mass $m_e$. \\
Also, the electron plasma-$\beta$ is defined by $\beta_e=\mu_0\,k_{\rm B}\,T_{0,e}\,n_{0} / B_{0,z}^2$, whereas
$\beta_i=\beta_e\tau$ for ions. 
\subsection{The gyrokinetic model}
The particle-in-cell code EUTERPE \cite{THP3-06} uses a $\delta f$-scheme splitting of the distribution function $f_s$
for each species $s$ into a time independent background $f_{0,s}$ and a perturbed part $\delta f_s$
in order to solve the full standard gyrokinetic Vlasov-Maxwell-system \cite{Hahm_2009} 
globally in toroidal 3D-geometry.
Here the code is modified to simulate the tearing mode in slab geometry.
The background distribution function is assumed to be a shifted Maxwellian with bulk velocity $u_{0,s}$.
 EUTERPE works in the $p_{\|}$-formalism so that the equations for particles trajectories are 
in a slab geometry \cite{Hahm_1988,Hahm_2009}
\begin{eqnarray}
\dot{\vec{R}}_s &=& \frac{1}{\alpha\,\sqrt{\beta_e}} \frac{p_{\|}}{m_s} \vec{b} -\Omega_s\,A\, \vec{b} + 
\,\vec{b}\, \times \, \nabla \langle  \Psi \rangle \\
&=& \frac{1}{\alpha\,\sqrt{\beta_e}} \frac{p_{\|}}{m_s} \vec{b} + \dot{\vec{R}}_s^1 \\ 
%$EU
%\dot{p}_{\|,s} &=& -\frac{q_s}{m_s} \vec{b} \, \nabla \langle  \phi-p_{\|}\,A \rangle
\frac{\dot{p}_{\|,s}}{m_s} &=& -\alpha\,\sqrt{\beta_e}\,\Omega_s \, \vec{b} \cdot \nabla \langle  \Psi \rangle,
\label{eq_gk1}
\end{eqnarray}
where $\Omega_s=\left(q_s\,B_{0,z}/m_s\right)L/v_{\rm A}$ is the cyclotron frequency normalized to the Alfv{\'e}n 
time for each species, $q_s$ is the species charge, $\Psi=\Phi - A \, p_{\|}/m_s$,
$\vec{b}$ is the normalized magnetic field and $\alpha=L/\rho_{{\rm S},e}$.
The perturbed distribution function is pushed along the particle orbits
according to
\begin{eqnarray}
\dot{\delta f_s} &=& -\dot{f}_{0,s} \label{eq_gk2} \\
&=& -f_{0,s} \left( \kappa_{u_{0,s}}\, \dot{\vec{R}}_s^1 \cdot \nabla x +
\frac{\Omega_s\,\sqrt{\beta_e}\,\alpha}{v_s^2} \left( \frac{p_{\|}}{m_s} -u_{0,s} \right) 
\vec{b} \cdot \nabla \langle \Psi \rangle \right). %\nonumber
\end{eqnarray}
The current gradient term with the bulk velocity $u_{0,s}$ reads
\begin{equation*}
\kappa_{u_{0,s}}(x)=\frac{\frac{p_{\|}}{m_s}-u_{0,s}}{v_{s}^2} \, \frac{\textrm{d} u_{0,s}}{\textrm{d} x} 
\end{equation*}
and $v_{s}=\sqrt{k_{\rm B}\,T_s/m_s}/\left( v_{\rm A}\,\hat{d}_i/L \right)$ is the normalized thermal speed of each species. 
The quasineutrality condition for drift kinetic electrons and gyrokinetic ions reads
\begin{equation}
n_e = \langle n_i \rangle + \frac{\Gamma_0 - 1}{\rho_i^2}\, \Phi.
\label{eq_fullpolar}
\end{equation}
$\Gamma_0$ is an integral operator that describes the average of the electrostatic potential
over a gyro-ring around the guiding center position.
If necessary the polarization density is approximated by a Pad{\'e} approximation 
due to the relative complex structure of $\Gamma_0$ in real space.
Otherwise
%}
the ion response is simplifed by using a long wavelength approximation, $k_{\perp}^2\rho_i^2 \ll 1$. Expanding 
$\Gamma_0-1$ in a Taylor series in this limit, the quasineutrality condition becomes
\begin{equation}
n_e = \langle n_i \rangle + \nabla_{\perp}^2\, \Phi \, .
\label{eq_qn_lwa}
\end{equation}
The gyroaveraging of the ion guiding center density perturbation, $n_i$, can be expressed by the phase space
integral 
\begin{eqnarray}
\langle n_i \rangle(\vec{x}) &=&
\int\,J\,\textrm{d}^6Z\,\delta\left(\vec{R}+\vec{\rho}_i-\vec{x}\right) \delta\,f_i
\end{eqnarray}
with the phase space Jacobian $J=B$, 
${\rm d}^6Z={\rm d}\vec{R}\,({\rm d}p_{\|}/m_i)\,v_{\perp}{\rm d}v_{\perp}{\rm d}\alpha$ 
and the gyroradius vector $\vec{\rho}_i(\alpha)$. \\
Amp{\`e}re's law closes the Vlasov-Maxwell-system
\begin{equation}
-\nabla_{\perp}^2 A + \sum_s \frac{\beta_s}{\rho_s^2}  A  = \sum_s  \langle j_{\|,s} \rangle
\label{eq_ampere_w_skin}
\end{equation}
with the Larmor radii $\rho_s=\sqrt{m_s\,k_{\rm B}\,T_s}/ \left(e\,B_{0,z}\right)$ for each species. 
The 
corresponding gyroaveraged ''current'' response is calculated according to
\begin{eqnarray}
\langle j_{\|,i} \rangle(\vec{x}) &=&
\int\,J\,\textrm{d}^6Z\,\delta\left(\vec{R}+\vec{\rho}_i-\vec{x}\right) \frac{p_{\|}}{m_i} \delta\,f_i.
\end{eqnarray}
\\
The gyroaverging procedure of the fields $A,\,\Phi$ is being employed according to
\begin{eqnarray}
\langle A,\Phi \rangle(\vec{R}) = \frac{1}{2\,\pi}\,\int \,\left( A,\, \Phi \right) ( \vec{R} +\vec{\rho} ) \, \textrm{d}\alpha,
\end{eqnarray}
taking sufficiently many points on the gyro ring around the guiding center position $\vec{R}$.
Recently, serious computational difficulties concerning the skin
terms in Eq.~(\ref{eq_ampere_w_skin}) could be resolved using an enhanced control variate method \cite{Hatzky_2007}.
\subsection{The gyrofluid model}

We consider the gyrofluid model that has been adopted in Refs.~\cite{CGTW_2012,CWG_2012,CGWB_2013} to investigate 
magnetic reconnection in collisionless high-temperature plasmas with a strong guide field. 
This model is obtained from 
the equations of Ref.~\cite{WT_2012} by neglecting magnetic curvature effects and assuming two-dimensional dynamics with 
$\partial/\partial z=0$, being $z$ the direction of the strong guide field. In turn, the model of Ref.~\cite{WT_2012} was 
obtained from the equations of Ref.~\cite{SH_2001} by taking only the first two velocity space moments of the gyrokinetic 
equations for both the electrons and the ions, assuming constant temperatures and neglecting collisions and the electron 
gyroradius. Electron inertia terms, on the other hand, were retained in order to break the frozen-in condition and allow 
for magnetic reconnection phenomena. Therefore, the evolution equations of this gyrofluid model consist of the continuity 
equation and the $z$-component of the equation of motion for the ion guiding centers:
\begin{equation}
\frac{\partial {n}_i}{\partial t}+ {[}\Gamma_0^{{1}/{2}} \Phi, {n}_i{]} = {[}{u}_i, \Gamma_0^{{1}/{2}} A {]} ,\label{e1}
\end{equation}
\begin{equation}
\frac{\partial D}{\partial t}+ {[}\Gamma_0^{{1}/{2}} \Phi, D{]} = \tau \rho_{{\rm S},e}^2 {[} \Gamma_0^{{1}/{2}} A, {n}_i {]} ,\label{e2}
\end{equation}
and similar equations for the electrons:
\begin{equation}
\frac{\partial n_e}{\partial t}+ {[}\Phi, n_e{]} = {[}u_e, A{]} ,\label{e3}
\end{equation}
\begin{equation}
\frac{\partial F}{\partial t}+ {[}\Phi, F{]} = - \rho_{{\rm S},e}^2 {[} A, n_e{]} ,\label{e4}
\end{equation}
where the Poisson brackets between two generic fields $f$ and $g$ are defined by 
$\left[f,g\right]=\vec{z} \cdot \nabla \, f \times \nabla\,g$. Here $D= \Gamma_0^{{1}/{2}} A + d_i^2 u_i$ is the
 ion guiding center parallel canonical momentum, whereas $F=A-d_e^2 u_e$ is the electron parallel 
canonical momentum. Furthermore, $\Gamma_0^{{1}/{2}} \Phi$ is the gyro-averaged electrostatic potential 
and $\Gamma_0^{{1}/{2}} A$ is the gyro-averaged parallel magnetic potential, where the symbol 
$\Gamma_0^{{1}/{2}}$ refers to the gyro-averaging operator that we adopt in its lowest-order Pad\'e approximant form~\cite{DH_1993}
\begin{equation}
\Gamma_0^{{1}/{2}} = \dfrac{1}{{1 - \dfrac{{\rho_i^2}}{2}\nabla_\bot^2}}.% ,
\end{equation}
This approximation gives reasonable values for the whole range of $k_\bot ^2 \rho_i ^2$.
The system of equations is completed by the parallel 
component of Amp\`ere's law, 
\begin{eqnarray}
\nabla_{\perp}^2 A&=&u_e-\Gamma_0^{{1}/{2}} u_i
\label{eq_gf_amp}
\end{eqnarray}
and by the quasineutrality condition
\begin{eqnarray}
n_e = \Gamma_0^{{1}/{2}} n_i  + \frac{\Gamma_0 - 1}{\rho_i^2}\, \Phi.
\label{eq_gf_qn}
\end{eqnarray}
The resulting model 
is dissipationless and suitable for the study of reconnection mediated by electron inertia. In particular, 
it possesses a noncanonical Hamiltonian structure~\cite{WT_2012} that reveals the presence of four Lagrangian invariants, 
which have proved to be helpful to understand how the reconnection evolution is affected by the plasma $\beta$ and by the 
ratio of species temperatures {~\cite{CGTW_2012,CGWB_2013}.}
\subsection{Equilibrium configuration and numerical setup}

To investigate spontaneous reconnection, the model equations are solved numerically 
with an initial equilibrium that is unstable with respect to tearing modes. 
The instability reconnects the antiparallel component of magnetic field lines at the resonant surface 
defined by $\vec{k} \cdot \vec{B}_{0}=0$, with
$\vec{k}$ indicating the wavevector of the mode. We consider a two-dimensional slab geometry with $x$ as the
coordinate of the equilibrium inhomogeneity~and~setting~$\partial/\partial z=0$.
The equilibrium magnetic field $\vec{B}_{0}$ results from an equilibrium current $u_{0,e}$ carried by electrons only
(for ions $u_{0,i}=0$). The perpendicular sheared magnetic field can be deduced from
a parallel vector potential ${A}_{0,\|}(x)$, which is chosen to be
\begin{eqnarray}
{A}_{0,\|}(x)&=&\frac{C}{\cosh^2\left({x}\right)}.
\label{eq_equilibrium}
\end{eqnarray}
The parameter $C$ was chosen to be $C=0.1$ if not stated otherwise. 
This results in a maximal relative shear strength of $B_{0,y}/B_{0,z} \sim 0.08$ in the domain and a shear length 
$l_s=B_{0,z}/(\textrm{d}B_{0,y}/\textrm{d}x)=5$ at the resonant surface $x=0$. \\
Furthermore, 
the plasma is considered homogeneous with flat density $n_{0,s}(x)=n_{\textrm{eq}}$ and temperature profiles 
$T_{0,s}(x)=T_{0,s}$ for every species $s$.
We considered a simulation domain $\{(x,y):-\pi \leq x \leq \pi, -a\pi \leq y \leq a\pi\}$, 
where the parameter $a$ fixes the domain length $L_y$ in $y$-direction, which is linked to the wavenumber $k_y=2\pi\,m/L_y$ of
the longest wavelength mode $m=1$ of the system. The tearing mode stability quantity~\cite{FKR_1963}  \dpr\, 
is then charaterized by the wavenumber
$k_y$ according to the analytical expression~\cite{Porc_2002}
\begin{eqnarray}
\Delta^{\prime}&=&2\,\frac{(3+k_y^2)(5-k_y^2)}{k_y^2 \sqrt{4+k_y^2}}.
\label{eq_dpr}
\end{eqnarray}
The tearing mode becomes unstable in nonideal MHD if $\Delta^{\prime} > 0$, which is the case if $k_y<\sqrt{5}$. \\
The field equations in EUTERPE are discretized in real space by a B-spline finite element method~\cite{Hatzky_2002}. 
The $y$-direction is treated periodically, while the fields $A$ and $\Phi$
are subject to Dirichlet boundary conditions with respect to $x$. 
For the simulations a resolution of up to $1024 \times 128$ grid points has been used for the $x$ and
$y$-direction, respectively. 
The code pushes the perturbed distribution function $\delta f_s$ 
along particles trajectory using a Runge-Kutta-scheme of fourth order. 
In the gyrokinetic simulations no special initial perturbations are chosen so that the tearing 
instability evolves out of noise. \\
The gyrofluid code decomposes the fields into a time-independent background equilibrium and an evolving
perturbation within a pseudospectral method~\cite{CGTW_2012}. Periodic boundary conditions are employed in both 
the $x$- and $y$-directions, and a grid of $1024 \times 128$ points has been used. Since periodic boundary conditions 
are imposed also along the $x$-direction, a Fourier series truncated to eleven modes is used to approximate 
Eq.~(\ref{eq_equilibrium}). Finally, an Adams-Bashforth algorithm is applied to push the fields in time, and an 
initial disturbance on the out-of-plane current density 
of width $\mathcal{O}(d_e)$ around the resonant surface is set to accelerate 
the onset of the tearing instability. \\
It is important to note that the boundary conditions for the fields with respect to the $x$-direction
are different in the two codes. 
This is a consequence of the historical development of the codes. Due to the numerical 
method underlying the gyrofluid code periodic boundary conditions arise naturally.
In EUTERPE the chosen field boundary conditions are fixed in the code.
Our choice of the domain size in the $x$-direction is sufficient to avoid finite domain size
effects on the value of the tearing stability index $\Delta^{\prime}$. However, 
in the following we will check the effects of the boundary conditions by performing a detailed linear
benchmark with an eigenvalue approach.
If, in the following, simulations in the drift kinetic limit were performed,
this was achieved by setting the temperature ratio to $\tau=1/900$, giving
$\rho_i = 1/30 \ll d_e$, which makes the effect of the gyroaveraging operators
negligible. Additionally, instead of the Pad\'e approximation the long
wavelength approximation was then used for the quasi-neutrality equation in
EUTERPE.

\section{Linear comparison of the models}  \label{sec3}
As a first step we check the accuracy of the codes in the linear regime with a benchmark. For this purpose 
a numerical eigenmode and eigenvalue analysis is applied to each of the two models in the drift
kinetic limit. After the 
accuracy of the codes is checked to a high degree, we proceed with a comparison of the models in both the 
drift kinetic limit and the finite Larmor radius case.
\subsection{Eigenvalue equations}
In this section we describe the procedure of performing a numerical
benchmark using a shooting method to get the linear dispersion relation in the
drift kinetic limit. 
An analysis of the eigenvalues and the eigenmode structure is given here for both
the linearised gyrofluid and the gyrokinetic equations.
The gyrofluid equations~(\ref{e1}--\ref{eq_gf_qn}), and the gyrokinetic equations~(\ref{eq_gk1}--\ref{eq_qn_lwa}), 
are linearised using the ansatz 
$\delta \Phi\left(x,y,t\right)={\rm e}^{i\left( k_y\,y-\omega t\right)} \tilde\Phi(x)$
and $\delta A\left(x,y,t\right)={\rm e}^{i\left( k_y\,y-\omega t\right)} \tilde A(x)$ for the perturbed quantities,
additionally assuming a long-wave-length approximation for the quasineutrality equation, Eq.~(\ref{eq_qn_lwa}).
The field equations
are cast into a general form with the coefficients $q^i_j$, with $(i,j)=(A,\Phi)$,
\begin{eqnarray}
\frac{\textrm{d}^2\tilde\Phi}{\textrm{d}x^2}&=&
-q^{\Phi}_{\Phi}\left(x,\omega \right)\tilde\Phi - q^{\Phi}_{A}\left(x,\omega \right) \tilde A \label{eq_evprob0} \\
\frac{\textrm{d}^2\tilde A}{\textrm{d}x^2}&=&-q^A_{\Phi}\left(x,\omega \right)\tilde\Phi - q^A_{A}\left(x,\omega \right) \tilde A
\label{eq_evprob}
\end{eqnarray}
The linearisation of the gyrofluid system gives the following coefficients
\begin{eqnarray}
q^{\Phi}_{\Phi}\left(x,\omega \right)&=&-k_y^2+\sum_s q_s \frac{F^{\prime}_{0,s}}{N_s}\,\frac{k_y}{\omega} \label{eq_fluidkoeffs0} \\
q^{\Phi}_{A}\left(x,\omega \right)&=& \sum_s \frac{q_s}{N_s} \left(-q_s-\frac{k_s}{k_{\|}} N_s-
\tau_s \rho_{{\rm S},e}^2 \frac{k_{\|} k_s}{\omega^2} \right)\\
q^{A}_{\Phi}\left(x,\omega \right)&=&\sum_s  -\frac{F^{\prime}_{0,s}}{N_s}\,\frac{k_y}{\omega}\\
q^{A}_{A}\left(x,\omega \right)&=&\sum_s \frac{q_s}{N_s} \left(-q_s-
\tau_s \rho_{{\rm S},e}^2 \frac{k_{\|} k_s}{\omega^2} \right)
\label{eq_fluidkoeffs}
\end{eqnarray}
where the prime denotes the derivative with respect to $x$. Also the quantities \\
\begin{eqnarray}
F^{\prime}_{0,s}&=&-B_{y,0}+\left(-1 \right)^{s+1} d_s^2\, u^{\prime}_{0,s} \\
k_{\|}&=&-A_0^{\prime} k_y \\
k_s&=&-u_{0,s}^{\prime} k_y \\
N_s&=&d_s^2\left( 1-\tau_s \frac{\rho_{{\rm S},e}^2}{d_s^2}  \frac{k_{\|}^2}{\omega^2} \right),
\end{eqnarray}
have been introduced to make the notation more compact. 
Note that in the above relations $\tau_1=\tau$ for ions and $\tau_2=1$ for electrons. \\
The coefficients resulting from the linearisation of the gyrokinetic model are
\begin{eqnarray}
q^{\Phi}_{\Phi}\left(x,\omega \right)&=&-k_y^2+\alpha^{2} \beta_{e} \sum_s \frac{\Omega_s}{\mu_s}
\, \frac{X_s}{v_s^2 \,k_{\|}} \langle V_s^1 \rangle \label{eq_kincoeffs0}\\
q^{\Phi}_{A}\left(x,\omega \right)&=&-\alpha \sqrt{\beta_e} \sum_s
\frac{\Omega_{s}}{\mu_{s}} \frac{X_s}{v_s^2 \,k_{\|}}\, \left(\langle V_s^2\rangle +
u_{0,s}\langle \,V_s^1 \rangle \right) \\
q^{A}_{\Phi}\left(x,\omega \right)&=&n_{e} \sum_s \frac{\Omega_s}{\mu_{s}} \,
\frac{X_s}{v_s^2 \,k_{\|}}
\left( \langle V_s^2\rangle + u_{0,s} \langle V_s^1\rangle \right) \\
q^{A}_{A}\left(x,\omega \right)&=&-k_y^2-\frac{n_{e}}{\alpha \sqrt{\beta_{e}}} \sum_s
\frac{\Omega_s}{\mu_{s}} \\
&\times& \left( \frac{X_s}{v_s^2 \,k_{\|}} \left( \langle V_s^3\rangle + 2 \, u_{0,s} \langle V_s^2\rangle 
+ u_{0,s}^2 \langle V_s^1 \rangle \right) + \Omega_s \right). \nonumber
\label{eq_kincoeffs}
\end{eqnarray}
where we have introduced $X_s=-k_s -k_{\|} \Omega_s$ and $\mu_{i}=1, \mu_{e}=\mu$. \\
The functions $\langle V_s^n \rangle\left(x,\omega\right)$ for each species are defined as
\begin{eqnarray}
\begin{split}
\langle V_s^n \rangle &= \left( \sqrt{2} \, v_s \right)^{n-1} \frac{1}{\sqrt{\pi}} 
\int^{\infty}_{-\infty} \textrm{d}t \, t^n \,  
\frac{{\rm e}^{-t^2}}{t-\left(\frac{1}{\sqrt{2}\,v_s}\left(\frac{\omega}{k_{\|}}-u_{0,s}\right) \right)} \\
&= \left( \sqrt{2} \, v_s \right)^{n-1} Z_n\left( \zeta_s \right)
\end{split}
\label{eq_pdf}
\end{eqnarray}
{with}  $Z_n\left( \zeta_s\right)$ being the plasma dispersion function of $n$-th order with the species
argument $\zeta_s=\left(\frac{\omega}{k_{\|}}-u_{0,s}\right)/\left(\sqrt{2}\,v_s\right)$. \\
These fourth-order equations are
 a nontrivial extension with respect to the case where the electrostatic 
potential $\tilde \Phi$ is negleted~\cite{Kat_1980, WCP_2005}, which is only of second order. 
Both these sets of eigenvalue equations are solved numerically using a shooting
method, which is formulated as a Riccati problem~\cite{mscott_1973}. 
By using an adaptive stepsize integrator results of very high accuracy results are obtained. \\
For the equilibrium configuration considered here,~i.\,e.~without
 any equilibrium gradients of temperature or density, the eigenvalue has only 
an imaginary part $\omega=i \, \gamma$. 
The Eqs.~(\ref{eq_evprob0}--\ref{eq_evprob}), with the coefficients 
(\ref{eq_fluidkoeffs0}--\ref{eq_fluidkoeffs}) and (\ref{eq_kincoeffs0}--\ref{eq_kincoeffs}),
are solved using Dirichlet boundary conditions in $x$-direction.\\
\subsection{Linear Benchmark with eigenvalue approach}
\begin{figure}
\centering
\begin{minipage}[hbt]{5cm}
	\centering
	\includegraphics[width=5cm,angle=-90]{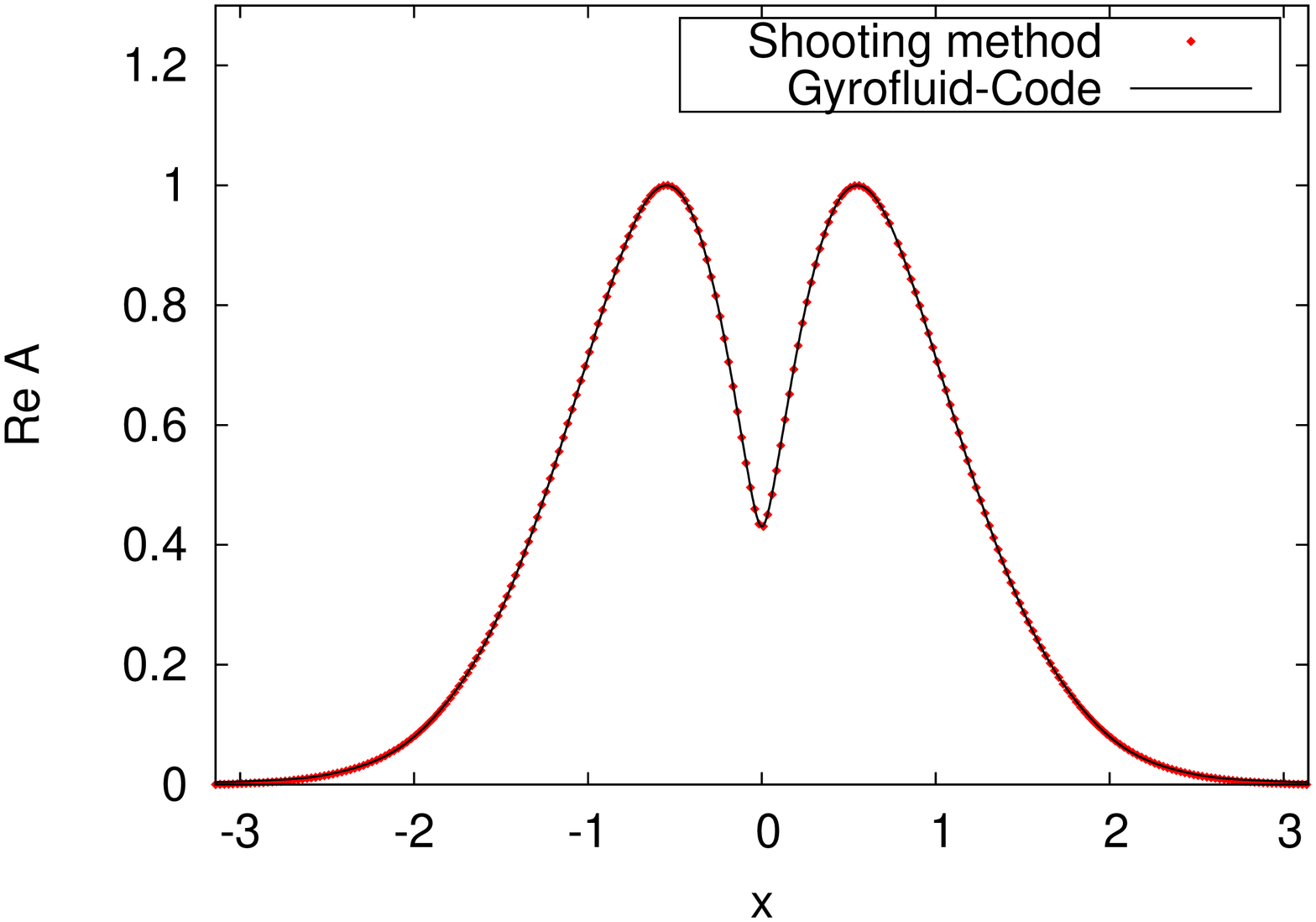}
	%\caption{Bild1}
	%\label{Bild1}
\end{minipage}
\hspace{3cm}%\hfill
\begin{minipage}[hbt]{5cm}
	\centering
	\includegraphics[width=5cm,angle=-90]{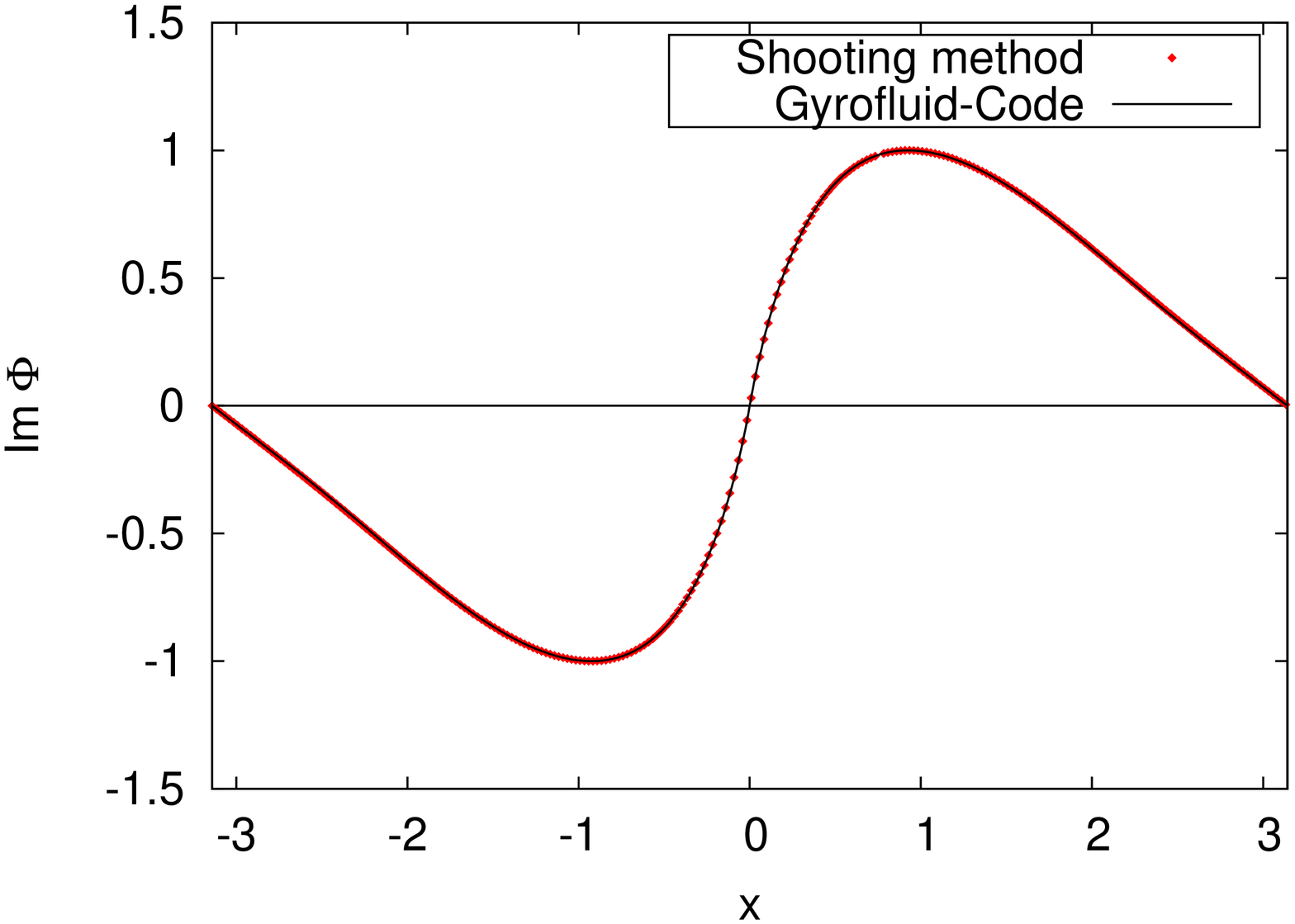}
	%\caption{Bild2}
	%\label{Bild2}
\end{minipage}
\caption{Benchmark of the gyrofluid eigenfunctions. Left the real part of the parallel 
vector potential $\tilde A$, right the imaginary part of the electrostatic potential $\tilde \Phi$. 
Eigenfunctions are normalized by their maximum value.}
\label{fig_ef_lu1}
\end{figure}

\begin{figure}
\centering
\begin{minipage}[hbt]{5cm}
	\centering
	\includegraphics[width=5cm,angle=-90]{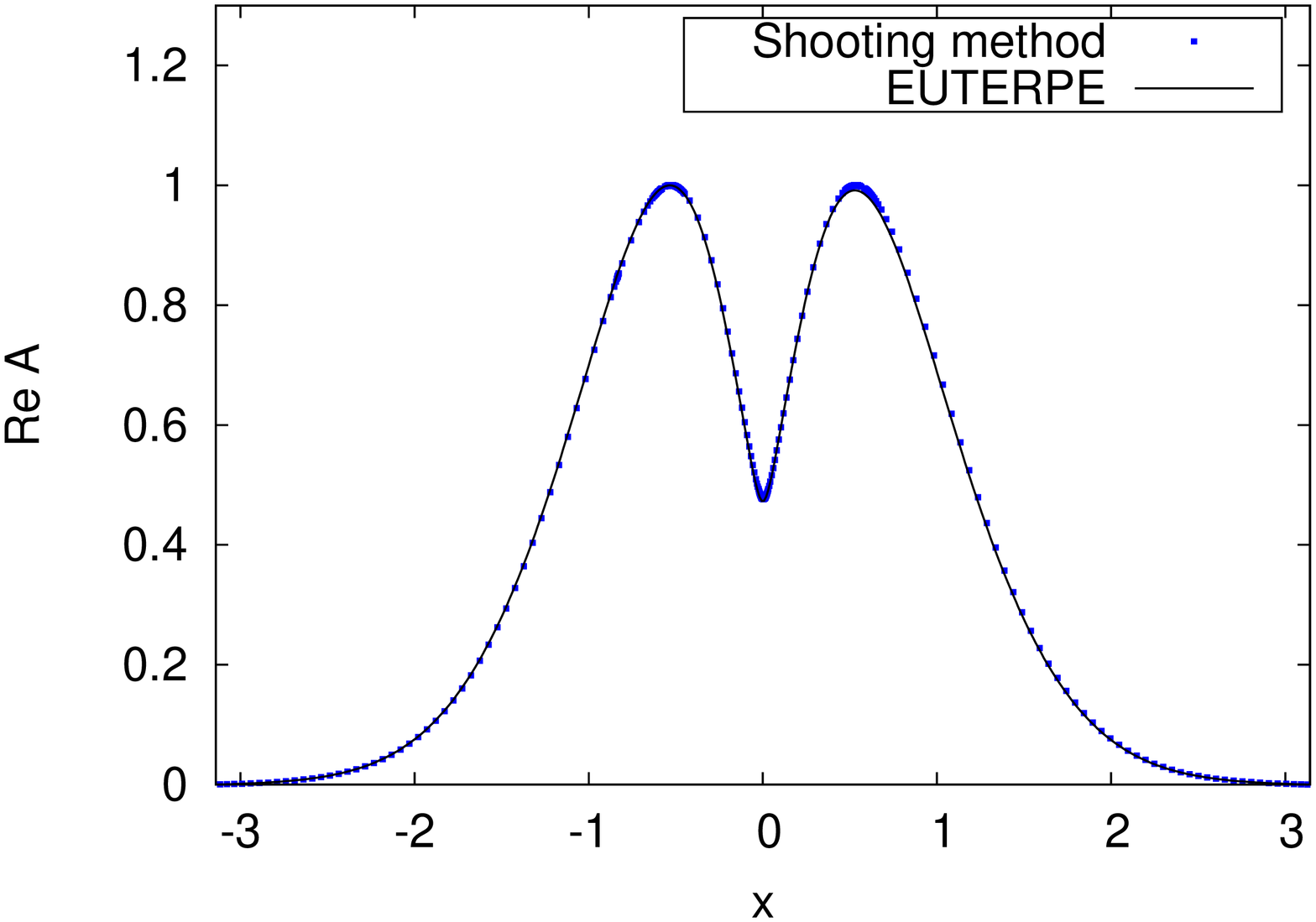}
	%\caption{Bild1}
	%\label{Bild1}
\end{minipage}
\hspace{3cm}%\hfill
\begin{minipage}[hbt]{5cm}
	\centering
	\includegraphics[width=5cm,angle=-90]{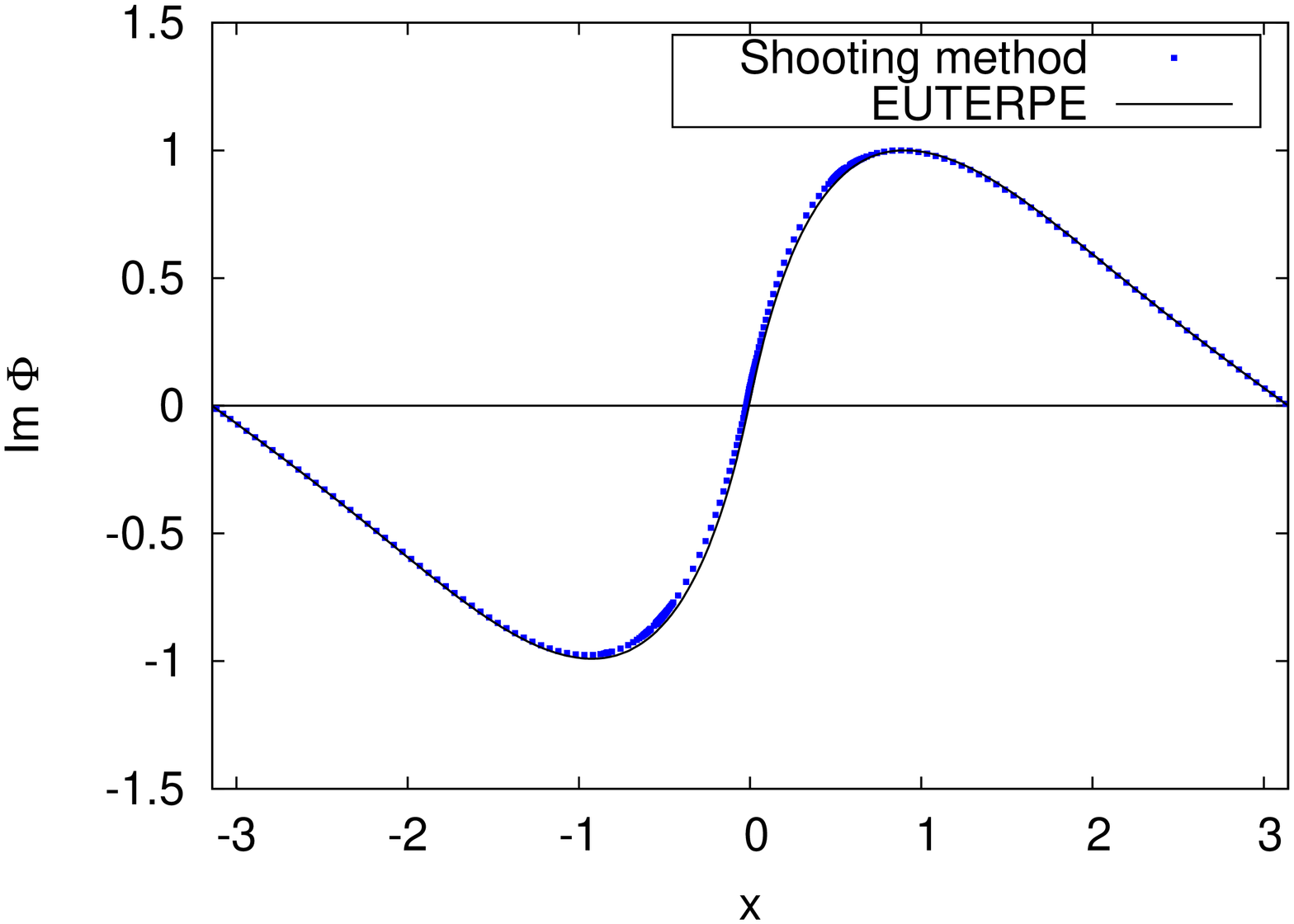}
	%\caption{Bild2}
	%\label{Bild2}
\end{minipage}
\caption{Benchmark of the gyrokinetic eigenfunctions. Left the real part of the parallel 
vector potential $\tilde A$, right the imaginary part of the electrostatic potential $\tilde\Phi$. 
Eigenfunctions are normalized by their maximum value.}
\label{fig_ef_eu1}
\end{figure}

\begin{figure}
  \centering
  \includegraphics[width=6.0cm,angle=-90,clip]{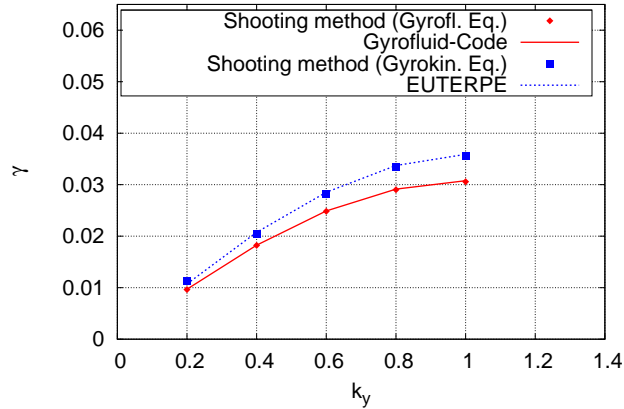}
  \caption{A benchmark of the linear growth rates of both models for various wavevectors $k_y$.
Both the gyrofluid code and gyrokinetic code work linearly exact.}
\label{fig_fullbench}
\end{figure}

The first benchmark is carried out for the parameter values $d_e=0.1, \,
d_i=4.285, \, \rho_{{\rm S},e}=0.6, k_y=0.6$ using the drift kinetic limit. This
corresponds to $\beta_e=1.96 \cdot 10^{-2}$ and a realistic proton to electron
mass ratio $\mu=1836$.
The comparison of the eigenfunction resulting from the shooting method
 with results from the gyrofluid simulation is shown in Figure~\ref{fig_ef_lu1}. 
Due to symmetries of the equations and the pure imaginary eigenvalue,
$\gamma=0.0248$,
only the real part of $\tilde A$ remains, as well as only an imaginary part of $\tilde \Phi$. 
The field structures agree very well with results from the shooting code, although the boundary
conditions with respect to $x$ differ.
%
%Lucas EF
The same procedure has been performed with EUTERPE 
using the coefficients defined by Eqs.~(\ref{eq_kincoeffs0}--\ref{eq_kincoeffs}).
In this case $\gamma=0.0273$, and both potentials are in good agreement with the results from the shooting
method as well, as shown in Figure~\ref{fig_ef_eu1}. 
In this case both methods used the same
boundary conditions regarding the $x$-direction. {The comparison with the solution of the gyrofluid 
problem shows} that the instability is mainly influenced by the dynamics at the
resonant layer. The solutions drop very fast to zero approaching the boundaries and
therefore the influence of the boundary conditions is suppressed.
This will be important for further nonlinear comparisons. \\ 
To check the eigenvalues over an extended $k_y$-spectrum of unstable modes, simulations have been
performed with the previous setup varying the simulation domain size $L_y$. 
The comparison of both fluid and kinetic results and
the relevant results of the shooting method are shown in Figure~\ref{fig_fullbench}.
We have thus shown numerically that the two codes give exact results in the linear regime
over a wide range of $k_y$. \\
\subsection{Model comparison in the drift kinetic limit}
\begin{figure}[h]
\begin{minipage}[hbt]{5cm}
	\centering
	\includegraphics[width=5cm,angle=-90]{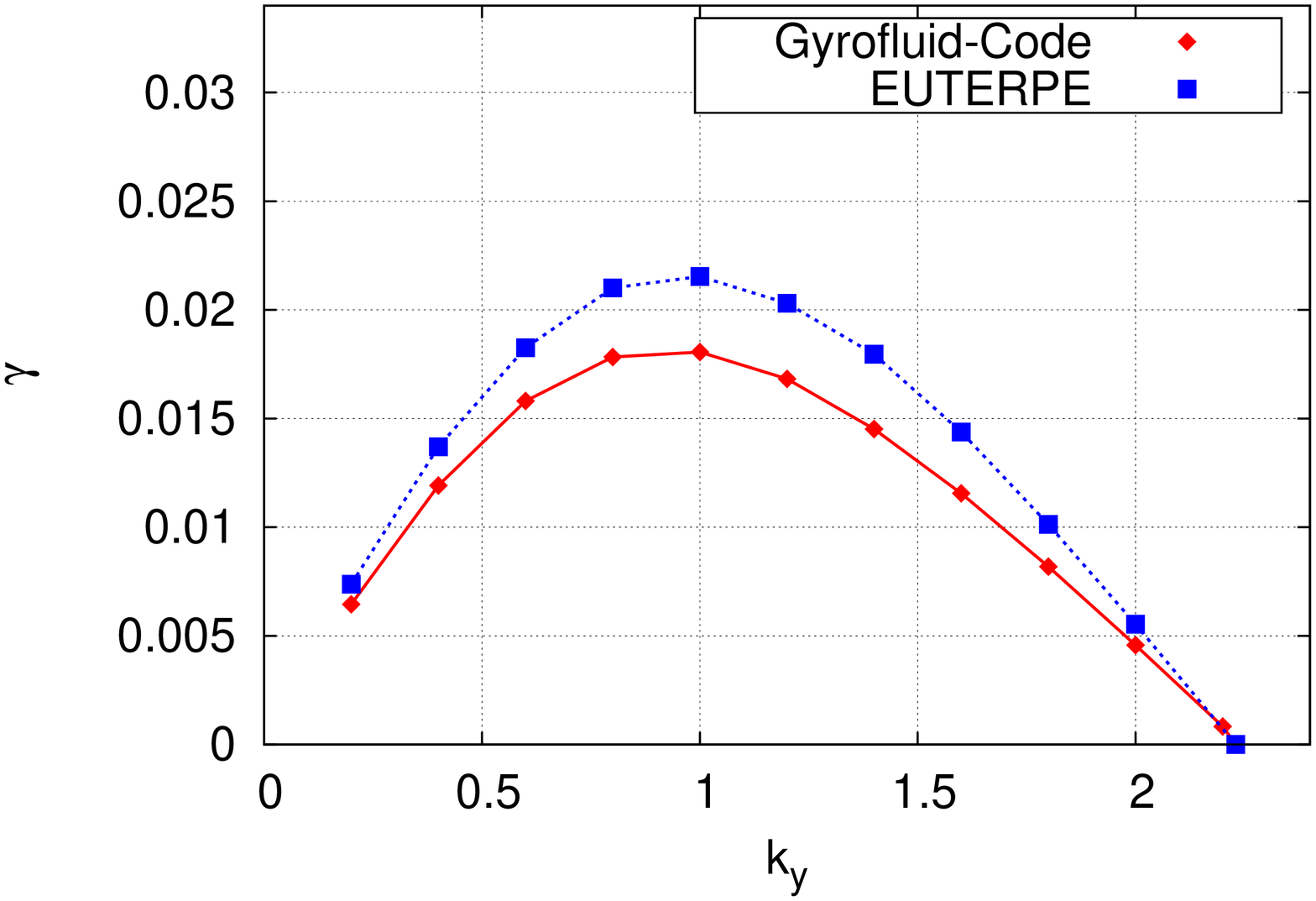}
	%\caption{Bild1}
	%\label{Bild1}
\end{minipage}
\hspace{3cm}%\hfill
\begin{minipage}[hbt]{5cm}
	\centering
	\includegraphics[width=5cm,angle=-90]{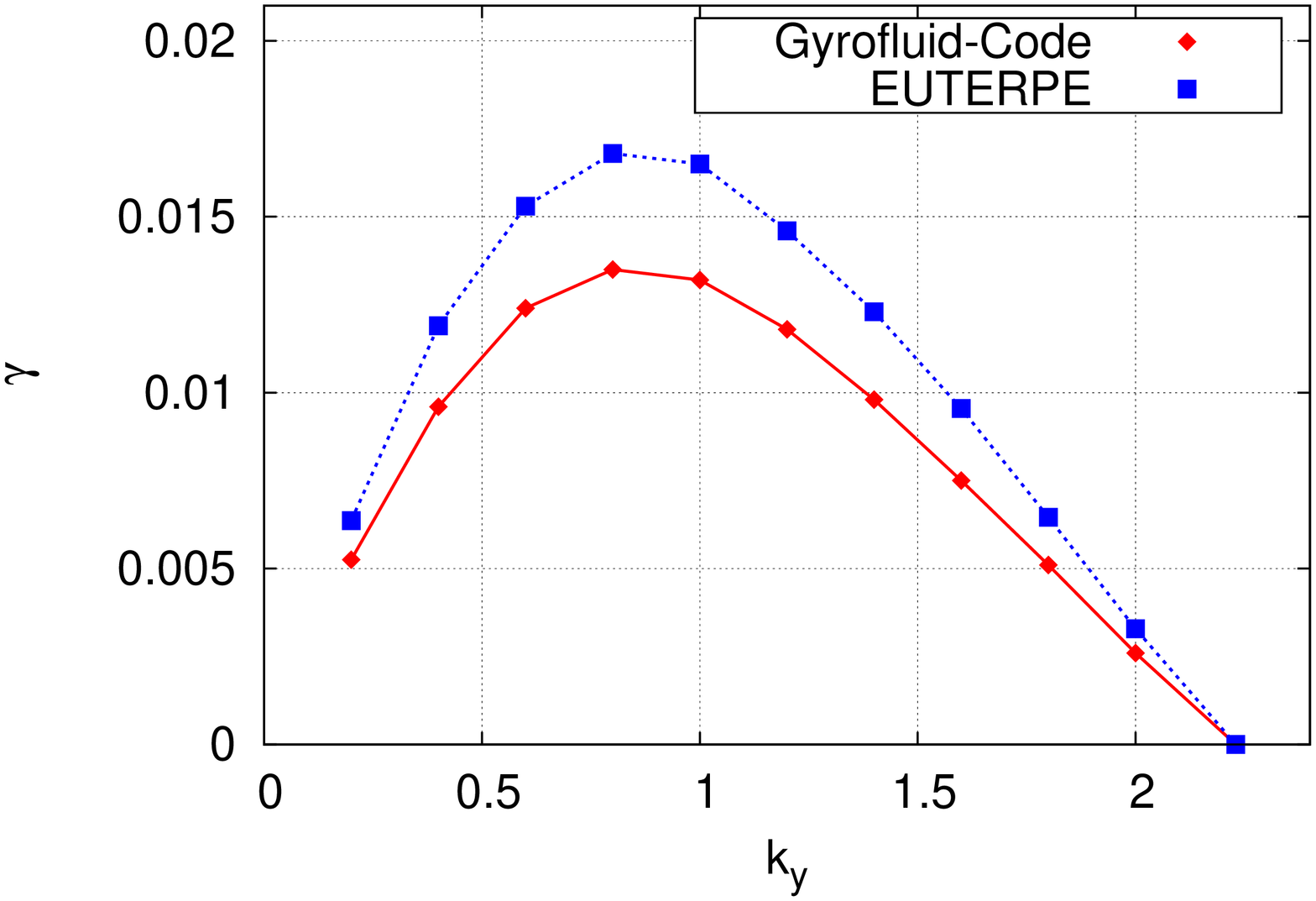}
	%\caption{Bild2}
	%\label{Bild2}
\end{minipage}
\caption{The comparison of the linear dispersions shows a good agreement between the two approaches 
over the full $k_y$ range. Left. Case I, $\mu=1836$. Right. Case II, $\mu=100$.
(Solid lines connect the numerical results for better visualization.)}
\label{fig_first_dr}
\end{figure}

In the following we use two sets of parameters which are relevant for reconnection
physics. The parameter associated with Setup I and II are listed in the Table below.
Case I refers to a realistic mass ratio $\mu$ and ''kinetic'' regime, $\beta_e \gg m_e/m_i$,
or equivalently $\rho_{{\rm S},e} \gg d_e$,
whereas case II defines a ''medium'' range between kinetic and inertial regime, 
$\beta_e \sim m_e/m_i$. \\
\begin{table}
\begin{center}
\begin{tabular}{c|c|c} \hline \hline
Setup & I & II \\\hline
$\mu$ & 1836 & 100  \\\hline
$\beta_e$ & $4.91\cdot 10^{-3}$ & $4 \cdot 10^{-2}$  \\\hline
$\rho_{{\rm S},e}$ & 0.3 & 0.2  \\\hline
$d_e$ & 0.1 & 0.1  \\\hline
$d_i$ & 4.285 & 1.0  \\\hline
\end{tabular}
\caption{Set of parameters defining setup I and II used for the simulations. }
\label{table1}
\end{center}
\end{table}
Simulations for cases I and II have been performed for various $k_y$. Over the full range of
wave numbers, from the large-$\Delta^{\prime}$ to the small-$\Delta^{\prime}$ cases, 
close to the stability threshold at $k_y \sim 2.23$,
both models describe the reconnection process very well, as shown in Figure~\ref{fig_first_dr}. 
It is found a relative maximum deviation of about $20\%$ around $k_y \sim 1$ for
both setups.
However, in the small-$\Delta^{\prime}$ limit the differences of the growth rates become smaller. \\
The kinetic description allows one to estimate the width of the region of particle
acceleration, $\delta$, due to the resonance condition 
$k_{\|}\,\rho_{{\rm S},e}/d_e=k_y\,\delta/l_s \cdot \rho_{{\rm S},e}/d_e\sim\gamma$ 
 in the small-$\Delta^{\prime}$ limit and $\delta \ll L$~\cite{DL_1977}. 
This limit is defined by the condition $\Delta^{\prime}\,d_e \ll \left( d_e/\rho_{{\rm S},e}\right)^{1/3}$.
Together with the kinetic dispersion relation in this limit, 
$\gamma=k_y\,d_e\,\rho_{{\rm S},e}\,\Delta^{\prime}/l_s$, one gets the estimate $\delta\sim\Delta^{\prime}\,d_e^2$. 
The two-fluid description also yields this scaling of the growth rate and current layer in the
small-$\Delta^{\prime}$ limit~\cite{Zocco_2011, P_1991}.\\
Another point which might be important concerns the assumptions of the 
adopted gyrofluid model, which is a truncation of the much more complete
model proposed by Snyder and Hammett~\cite{SH_2001}.
The derivation uses the restriction that the bulk velocity of the species $u_{0,s}$ is much smaller 
than the thermal velocity $v_s$. Moreover, this model uses an unshifted Maxwellian when performing
 the integration over the velocity space to get the equations of moments. Therefore, the 
gyrofluid equations hold exactly only for $C \ll 1$. 
For the linear simulations done here the amplitude of the sheared perpendicular field was chosen as 
$C=0.1$, which approximates this limit very well and additionally allows relatively short simulation times. 
However we checked this point simulating a mode with~$k_y=1.0, d_e=0.1, d_i=4.285, \rho_{{\rm S},e}=0.3$ and
decreasing $C$ from $10^{-1}$ to $10^{-4}$. Although these runs required very long simulation times
for small $C$, due to the dependence of $\gamma$ from~$l_s$, the relative deviation of the growth rates of the models fell from 
approximately~20$\%$ to 12$\%$.\\
\begin{comment}
It is important to note that the gyrokinetic equations (\ref{eq_gk2}) cover completely the initial gyrokinetic equation
of~\cite{SH_2001} in the linear case. Since the gyrokinetic model~\cite{SH_2001} contains second order terms with respect to the
 gyrokinetic ordering parameter $\epsilon$, discrepancies could maybe play a role in nonlinear situations.\\\\
\end{comment}
%
%
\subsection{Influence of gyro-effects}
\begin{figure}
\centering
\begin{minipage}[hbt]{5cm}
	\centering
	\includegraphics[width=5cm,angle=-90]{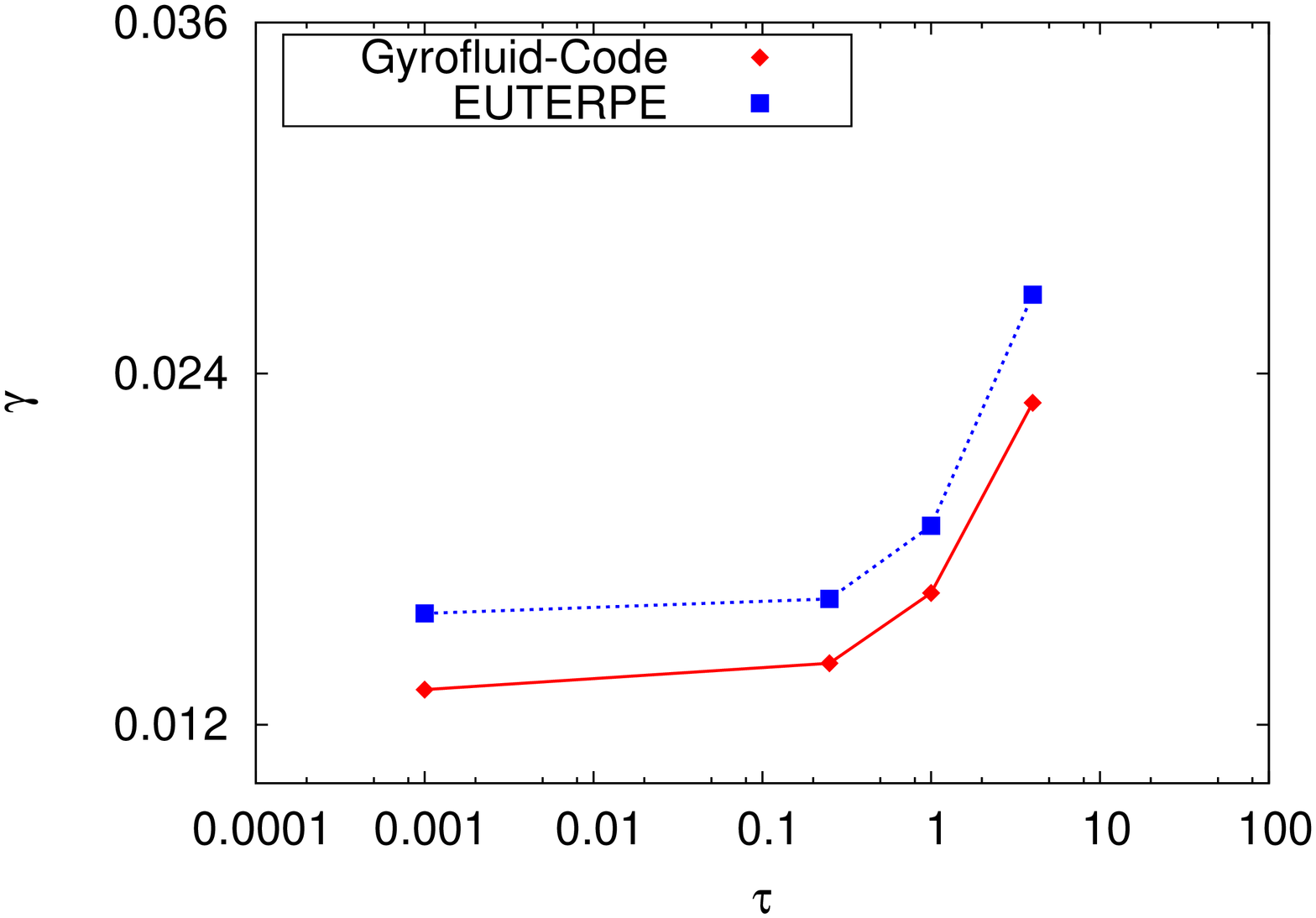}
%	\label{fig_padecyclone2}
\end{minipage}
\hspace{3cm}%\hfill
\begin{minipage}[hbt]{5cm}
	\centering
	\includegraphics[width=5cm,angle=-90]{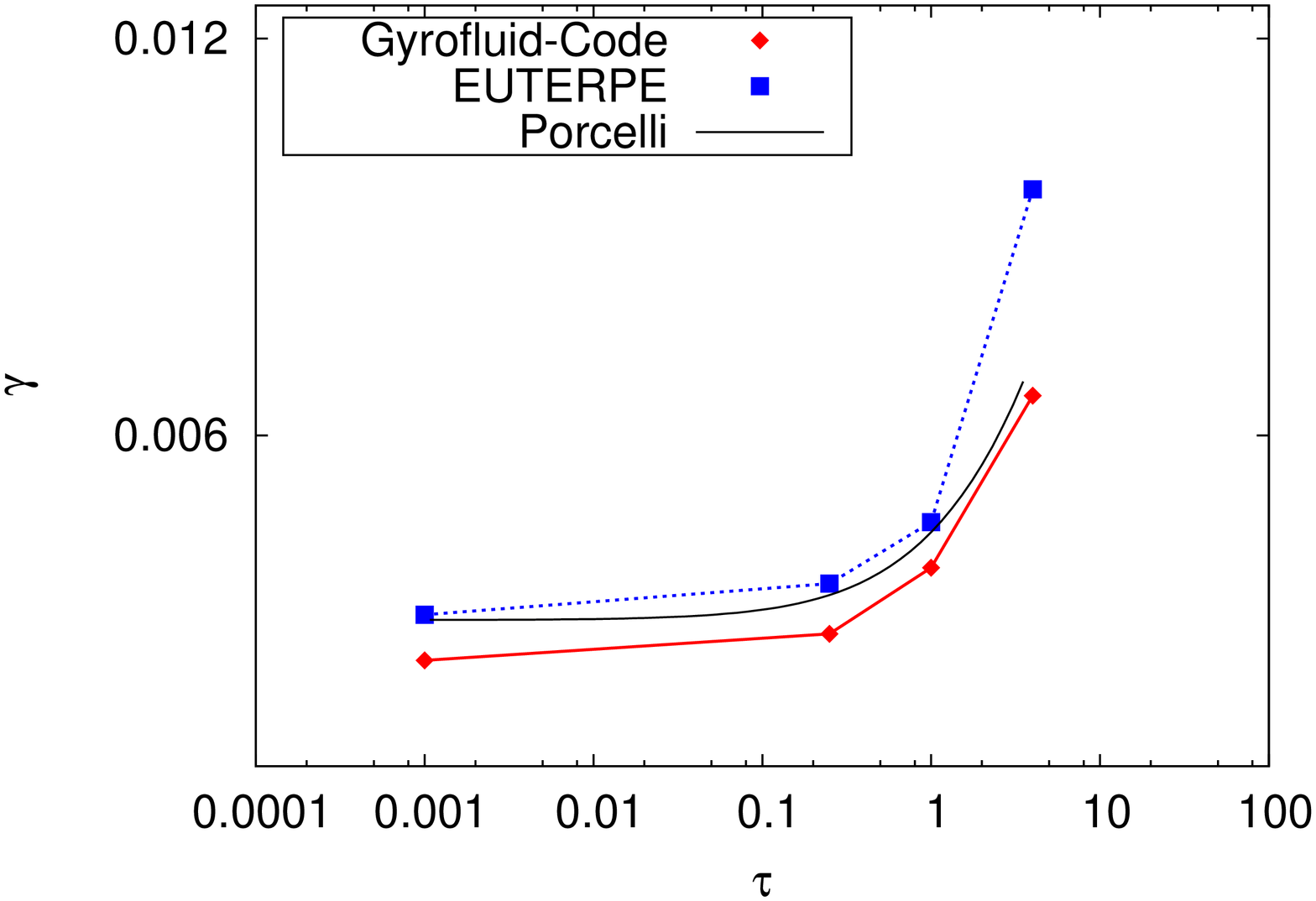}
%	\caption{}
%	\label{fig_padecyclone1}
\end{minipage}
	\caption{In the medium-$\Delta'$ regime, $\Delta^{\prime}\,d_e \sim 1$ (left), 
          and in the small $\Delta^{\prime}\,d_e$ regime (right) the codes show good 
          agreement over whole range of $\tau$. 
          The analytical predicition, Eq.~(\ref{eq_porc}), 
          fits well for both the gyrokinetic and gyrofluid model (right).}
\label{fig_padecyclone}
\end{figure}

It is desirable to go beyond the drift kinetic limit and simulate the tearing mode for
finite ion temperatures when the gyroradius can become much larger than the
thickness of the electron diffusion region which is $\mathcal{O}(d_e)$~\cite{Cowley_1985}.
Here we only compare the linear simulations of the codes using the setup scenario II
for $k_y=1.0$ and $2.0$, while varying $\tau$. 
The gyrokinetic effects now enter according to Eq.~(\ref{eq_fullpolar}) using the approximation of Pad{\'e}.\\ 
Figure~\ref{fig_padecyclone} shows that the growth rates obtained with the two
different codes behave qualitatively very similar when we vary
$\tau$. {While for small $\tau$ the growth rate remains nearly constant, for larger
ion-gyroradii $\rho_i \gg \rho_{{\rm S},e}$ ($\tau \gtrsim 1$), the growth
rate begins to increase strongly.}

 For the medium range $k_y \sim 1$ both models
cover the physics very well (Figure~\ref{fig_padecyclone}, left).  This result
is important since it proves clearly that the gyro-effects are being covered
correctly by both gyro-approaches, which provides a good starting point for the
following comparisons in the nonlinear regime. \\
The right frame of Figure~\ref{fig_padecyclone} displays the simulation
results in the small-$\Delta^{\prime}$
limit, which for the case with hot electrons and ions is defined by
$\Delta^{\prime}\,d_e \ll \left[d_e/\left( \rho_{{\rm S},e}\,\sqrt{1+\tau} \right)\right]^{1/3}$.
In this range of parameters an analytical prediction for a 
kinetic ion response together with an electron fluid derived by Porcelli gives~\cite{P_1991}
\begin{eqnarray}
\gamma&=&k_y\,\Delta^{\prime}\,\sqrt{1+\tau}\,\frac{d_e\,\rho_{{\rm S},e}}{l_s\,\pi},
\label{eq_porc}
\end{eqnarray}
which {reproduces} the simulation results to high accuracy. 
{Since the parallel ion dynamics and the gyrophase-independent part of the real space 
ion particle density $\langle n_i \rangle$ were neglected in Porcelli's theory, 
their effect plays a negligible role when considering an equilibrium without density gradients. 
Ion diamagnetic drifts may change this picture, and an investigation of nonuniform ion density equilibria 
will be the subject of a future publication. }\\

\section{Comparison of the nonlinear models} \label{sec4}

In fusion relevant applications the 
saturated behaviour of the tearing instability is a very important issue. 
Continuing with the parameters of 
both cases I and II we now discuss the nonlinear phase, concentrating on the 
small-$\Delta^{\prime}$ regime. 
The saturated island half width~$w$ and oscillation
frequency $\omega_{\rm B}$ in the deeply nonlinear phase are the two most relevant observables. 
Up to now, in the literature there are only a few
extended simulation results of these quantities in homogeneous plasmas~\cite{WCP_2005,Wang_2011, ZKH_2012}.\\
It is important to note that the 
{equilibrium considered in this section}
is unstable with respect to modes with $m=1$, 
which can in general interact in the nonlinear phase with the~$m=0$ mode. 
Pseudospectral codes simulate a complete rectangular
domain $\left[ -m_{\rm max}, \ldots, m_{\rm max}\right] \times \left[ -n_{\rm max}, \ldots, n_{\rm max} \right]$ 
in Fourier space~\cite{Lele_1992}, 
$n$ being the mode number in $z$-direction ($n_{\rm max}=0$ here), 
so the $m=0$ mode is being simulated as well. In the gyrofluid
simulations all relevant scales were well resolved by choosing the extent of the 
Fourier spectrum to $1/k_{\rm max} \ll d_e$. 
In EUTERPE it is not necessary to choose a corresponding domain setup. 
Nevertheless, to match the initial computational conditions of the two methods,
EUTERPE was adjusted to adopt the filter $\left[ -1, \ldots, 1\right] \times \left[ 0 \right]$. 
Because higher modes numbers $m=2,\,3,\ldots$ are expected to play no role in the
dynamics the chosen filter does not restrict the essential physics. \\
The gyrokinetic simulations were performed with up to $N_p=3\cdot 10^7$ markers 
with a minimum time step $\Delta t=0.125$.
The skin depth $d_e=0.1$ is resolved with at least 16 points, whereas the width of the perturbed 
current produced by the parallel electric field, $\delta$, was resolved with about ten points. 
The numerical resolution
of the vector potential in the $x$-direction amounts to $n_{\rm{x}}=1024$ points, which separates
scales up to $\Delta\,x=5 \cdot 10^{-3}$. This introduces an upper error range,
which can be removed with finer grid resolutions but demands a much higher computational effort. \\
We apply two different methods to obtain the island half widths $w$ of the collisionless
tearing mode.
Assuming the constant-$\tilde A$ approximation, the half width evolution is given by~\cite{GoRu_1995}
\begin{eqnarray}
  w(t)= 2 \, \sqrt{{\tilde A}(x=0,\,y=0,\,t) \, l_s}.
  \label{eq_constisle}
\end{eqnarray}
Otherwise, without any approximation, we can obtain the exact island half width 
using the geometric definition of the island separatrix 
at each time step by solving numerically the equation
\begin{eqnarray}
  A(x=0,y=0,t) = A\left(x=w\left( t\right),\,y=\frac{\pi}{k_y},\,t \right)
  \label{eq_geomisle}
\end{eqnarray}
on the discrete spatial grid used in the codes. 
Assuming that the $X$-point is at $x=0,\,y=0$ and following the separatrix, 
the island half width $w$ is found at $x=w\left( t \right),\,y=\pi/k_y$. 
\subsection{Drift kinetic limit}
\begin{figure}[ht]
\begin{center}
\includegraphics[width=6.0cm,angle=-90,clip]{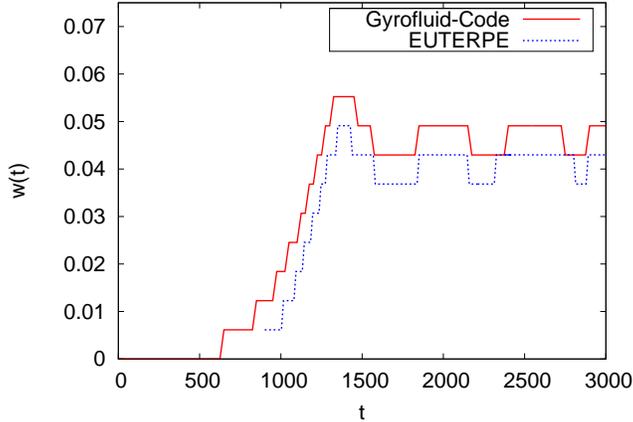}
\caption{Island half width as a function of time for setup I and $k_y=1.8$. 
Both gyrofluid and gyrokinetic models show clear saturated behaviour of the mode. The steps are due to
the spatial discrete grid points. }
\label{fig_nlin1}
\end{center}
\end{figure}

\begin{figure}[h]
\begin{center}
\includegraphics[width=6.0cm,angle=-90,clip]{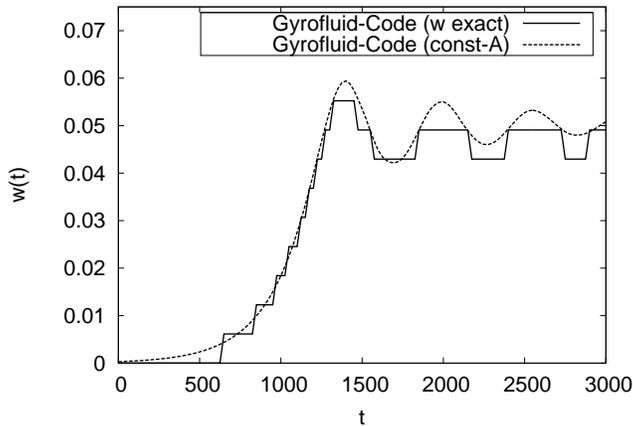}
\caption{Comparison between
the exact island half width obtained by solving Eq.~(\ref{eq_geomisle}) on a discrete 
spatial grid and the island half width calculated according to Eq.~(\ref{eq_constisle}) 
(Setup I and $k_y=1.8$).
In the small-$\Delta^{\prime}$ limit the const-$\tilde A$ approximation is numerically confirmed. }
\label{fig_constpsi_ky1.8_caseI}
\end{center}
\end{figure}

\begin{figure}[h]
\centering
\begin{minipage}[hbt]{5cm}
	\centering
	\includegraphics[width=5cm,angle=-90]{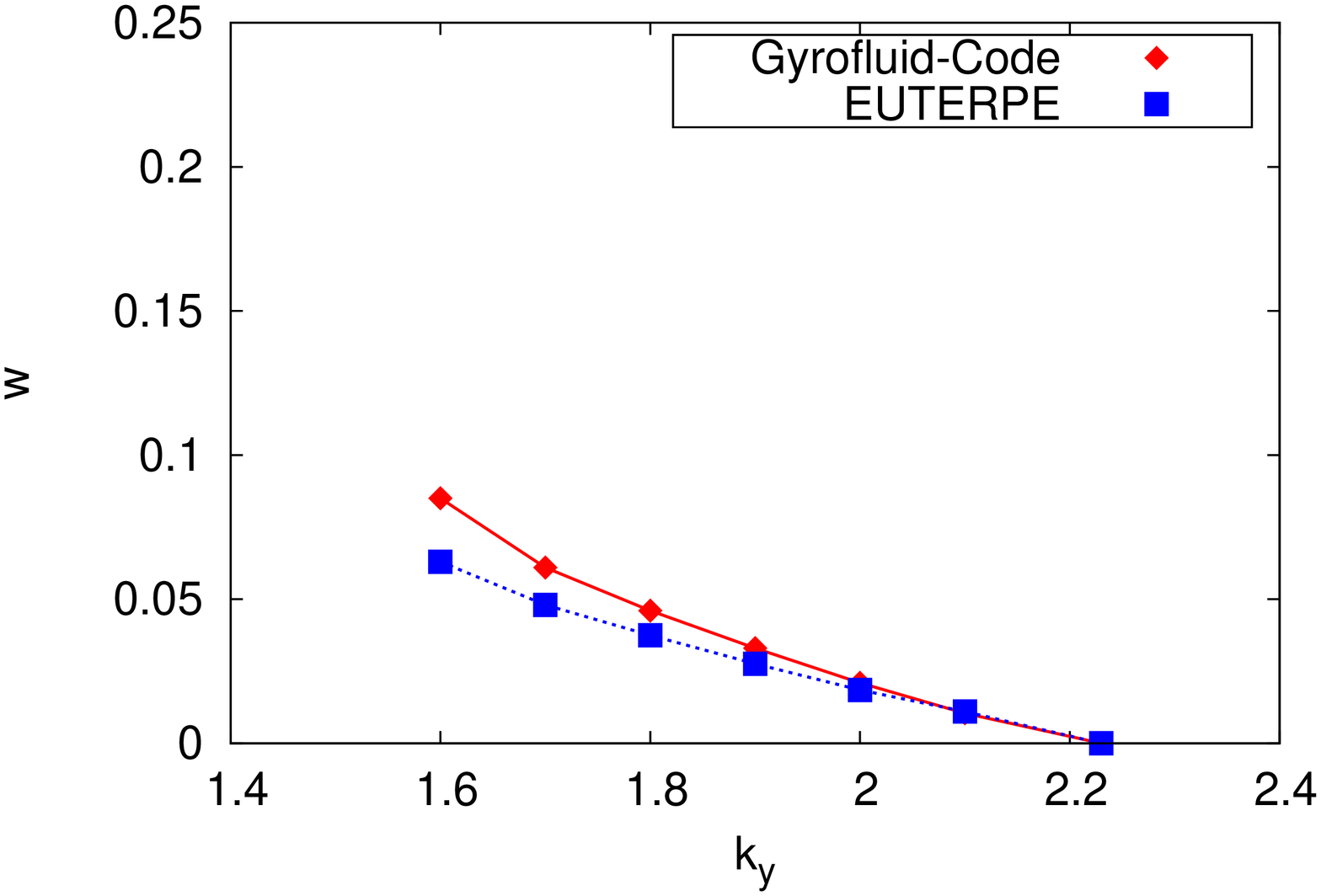}
	%\caption{Bild1}
	%\label{Bild1}
\end{minipage}
\hspace{3cm}%\hfill
\begin{minipage}[hbt]{5cm}
	\centering
	\includegraphics[width=5cm,angle=-90]{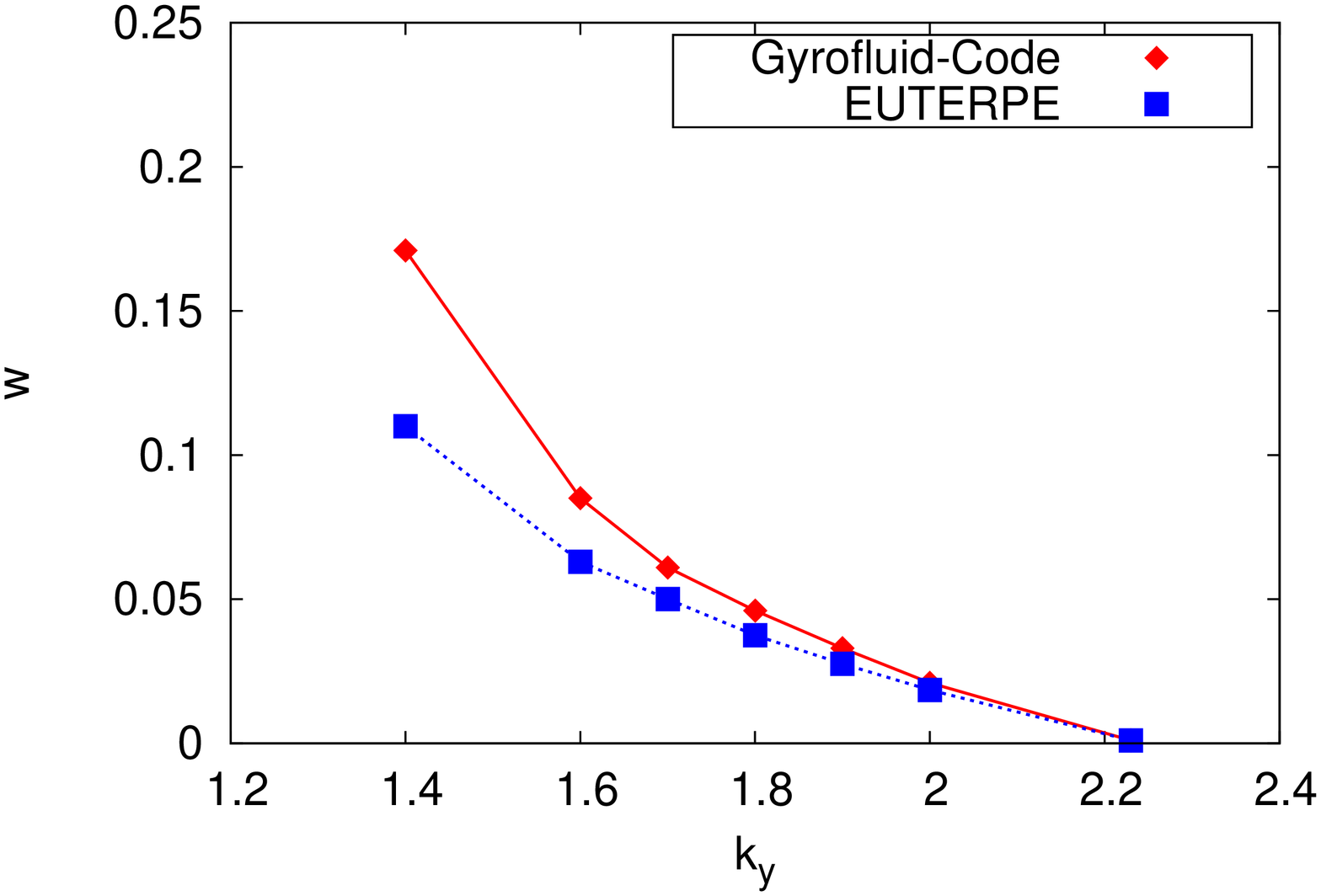}
	%\caption{Bild2}
	%\label{Bild2}
\end{minipage}
\caption{Saturated island half width $w$ as a function of $k_y$ for Setup I (left panel)
and for Setup II (right panel). The gyrokinetic and gyrofluid models show a very good
agreement in determining the saturated island half width in the small-$\Delta^{\prime}$ limit.}
\label{fig_w_bothcases}
\end{figure}

\begin{figure}[h]
  \centering
  \includegraphics[width=5cm,angle=-90]{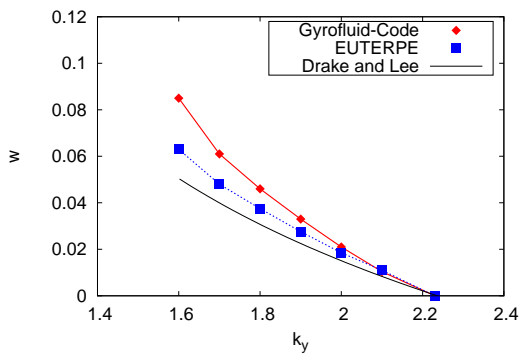}
  \caption{
The saturated island half width $w$ depending on $k_y$ is compared with the prediction by Drake and
Lee~\cite{DL_1977_PRL} for the parameter case I.
The analytical model shows a good qualitative agreement with simulation results for $\Delta^{\prime} < 1$. }
  \label{fig_nlinwithdrake}
\end{figure}

The evolution of the island half width into the deeply nonlinear regime 
is shown in Figure~\ref{fig_nlin1} for the parameters $k_y=1.8,\,\mu=1836$ 
and $\rho_{{\rm S},e}=0.3$ obtained with both codes. This Figure shows the
 solution of Eq.~(\ref{eq_geomisle}) at each time step.
Both gyrofluid and gyrokinetic models behave  well in the nonlinear phase and show a clear 
saturated phase beginning at~$t \sim 1500$.
The energy conservation proved to be more accurate than $2.5\%$.
Moreover, it turned out for all simulations presented here that the coupling between 
the $m=0$ and $m=1$ modes is very weak and can be neglected. 
Figure~\ref{fig_constpsi_ky1.8_caseI} shows a
comparison of the evolution of the exact island half width and the island half width obtained
according to Eq.~(\ref{eq_constisle}) for the gyrofluid simulation shown in Figure~\ref{fig_nlin1}.
We have checked that for wavenumbers $k_y \geq 1.8$, which corresponds to the small-$\Delta^{\prime}$
limit, 
the island half width calculated with the constant-$\tilde A$ approximation is valid within the
precision of measurement. Nevertheless, in the following we use Eq.~(\ref{eq_geomisle}).
\\
When the island width becomes comparable to the linear current sheet thickness $\delta$, the mode 
saturates~\cite{DL_1977_PRL}. After the transition into the saturation phase
the width of the island begins to oscillate with a
characteristic frequency $\omega_{\rm B}$, which is clearly visible in Figure 
~\ref{fig_nlin1} and~\ref{fig_constpsi_ky1.8_caseI}.
From the timeseries $w(t)$
the saturated island half width $w$ is measured by taking the mean value  $w=\langle w(t)\rangle_T$ 
after saturation starts, with $T$ indicating a period longer than the oscillation frequency. \\
In the following we measure both quantities $w$ and $\omega_{\rm B}$ 
for an extended parameter range to compare 
the gyrokinetic and gyrofluid models,
and to check the validity of analytical predictions in this regime of parameters. \\
Figure~\ref{fig_w_bothcases} shows {the saturated island half width} $w$ 
as a function of the longest wavelength in the system for both parameter cases. 
For values $k_y \sim 1.6$ the relative difference of the island half widths obtained
with the two adopted models is found to be about $30\%$
for both parameter cases I and II. 
Increasing $k_y$ to the range $k_y = 1.9 \ldots 2.23$
close to the
stability threshold the agreement between the results of 
the two codes is much better. The relative deviation of the island half widths
is approximately 10\% for $k_y=1.9$ in both setups and vanishes practically for higher wavenumbers.
This shows that for $\Delta^{\prime} \lesssim 1$ both models agree very well. 
Therefore, there are no significant differences between the 
gyrofluid and the gyrokinetic models for small island widths,~i.\,e.~when $w \lesssim d_e$. 
So for the cases investigated here, in which the island half width 
and the current layer thickness $\delta$  are much smaller than
the equilibrium scales, the fluid description produces practically the same 
island half widths as the more complete kinetic model. 
The comparison between the models also shows that the island width 
is slightly higher in the fluid description than in the kinetic model. 
These are the first extended comparisons of the saturated island width in slab geometry over
a broad range of parameter.
\\
Since for both parameter cases the ion skin depth 
is much larger than the electron skin depth, $d_e \ll d_i$, 
electron inertia dominates completely.
This regime has been investigated analytically
in an early kinetic approach by Drake and Lee~\cite{DL_1977_PRL}, where it was shown that the tearing mode saturates 
approximately when $w \sim \delta $, which in this regime means $w \sim \Delta^{\prime}\,d_e^2$. 
A more detailed estimation yields 
\begin{eqnarray}
  w&=&\Delta^{\prime}\,\frac{d_e^2}{2\,G}
  \label{eq_drakenlin}
\end{eqnarray}
with $G\sim 0.41$~\cite{DL_1977_PRL}. 
 Therefore, in the small island limit, $d_e \ll L \sim l_s$, 
the saturated island half width is described only by the skin depth $d_e$ and the 
tearing mode stability parameter $\Delta^{\prime}$, which for our choice of the 
equilibrium is known analytically from Eq.~(\ref{eq_dpr}). \\
The analytical prediction in comparison with our simulation results depending on $k_y$ is 
shown in Figure~\ref{fig_nlinwithdrake}.
Eq.~(\ref{eq_drakenlin}) well reflects the qualitative behaviour of $w$ over the shown
$k_y$-range, and agrees more closely with the gyrokinetic {results} than the gyrofluid ones. 
The deviations of the prediction of $w$ can be caused by assumptions which are not 
completely valid in the simulations. 
For instance, in the analytical estimations the shifted background 
Maxwellian was not used rigorously, and in addition the density response was neglected. \\
For both parameter cases investigated here, the island width 
does not seem to depend on the values of $\rho_{{\rm S},e}=0.2,\,0.3$,
as can be seen by comparing the left and right panels of Figure~\ref{fig_w_bothcases}. 
This suggests that there is no influence of finite electron temperature effects on the island width. 
This is consistent with the fact that
the analytical prediction, Eq.~(\ref{eq_drakenlin}), does not contain 
finite electron temperature effects related to $\rho_{{\rm S},e}$, which
are linked to finite pressure effects and the width of the ion inflow region~\cite{Zocco_2011}. 
Since $\rho_{{\rm S},e}$ is comparable to the electron skin depth and the analytical model does not
contain this quantity, 
it is unclear whether it plays an important role in nonlinear
simulations with both kinetic species. 
To investigate this dependence we fix the parameters $k_y=1.8,\, \Delta^{\prime}\,d_e \sim 0.25,\, \mu=1836$ and vary 
$\rho_{{\rm S},e}= 0.3,\,0.1,\,0.05,\,0.025$.
The simulations have shown that the island half width remains the same
$\left( w \sim 0.04 \right) $ to high accuracy in both
gyrokinetic and gyrofluid simulations. It follows that in the small-$\Delta^{\prime}$ regime the 
pressure scale has no influence on the saturation level of the collisionless tearing mode. \\\\
A further important nonlinear quantity which has been compared within 
the adopted gyrokinetic and gyrofluid models is the oscillation frequency $\omega_{\rm B}$ 
that characterizes the saturation phase, 
as shown in Figures~\ref{fig_nlin1} and~\ref{fig_constpsi_ky1.8_caseI}.
In the kinetic context it was observed that this frequency is due to the bounce motion 
of trapped electrons in the island~\cite{WCP_2005}.
We consider again the parameter cases I and II, and
measure the oscillation frequency as the mean value of several oscillation periods 
in the deeply nonlinear saturation phase, namely $\omega_{\rm B}=2\pi\,n_p/(T_f-T_i)$, being
$n_p$ the number of periods.
\begin{figure}[h]
\centering
\begin{minipage}[hbt]{5cm}
	\centering
	\includegraphics[width=5cm,angle=-90]{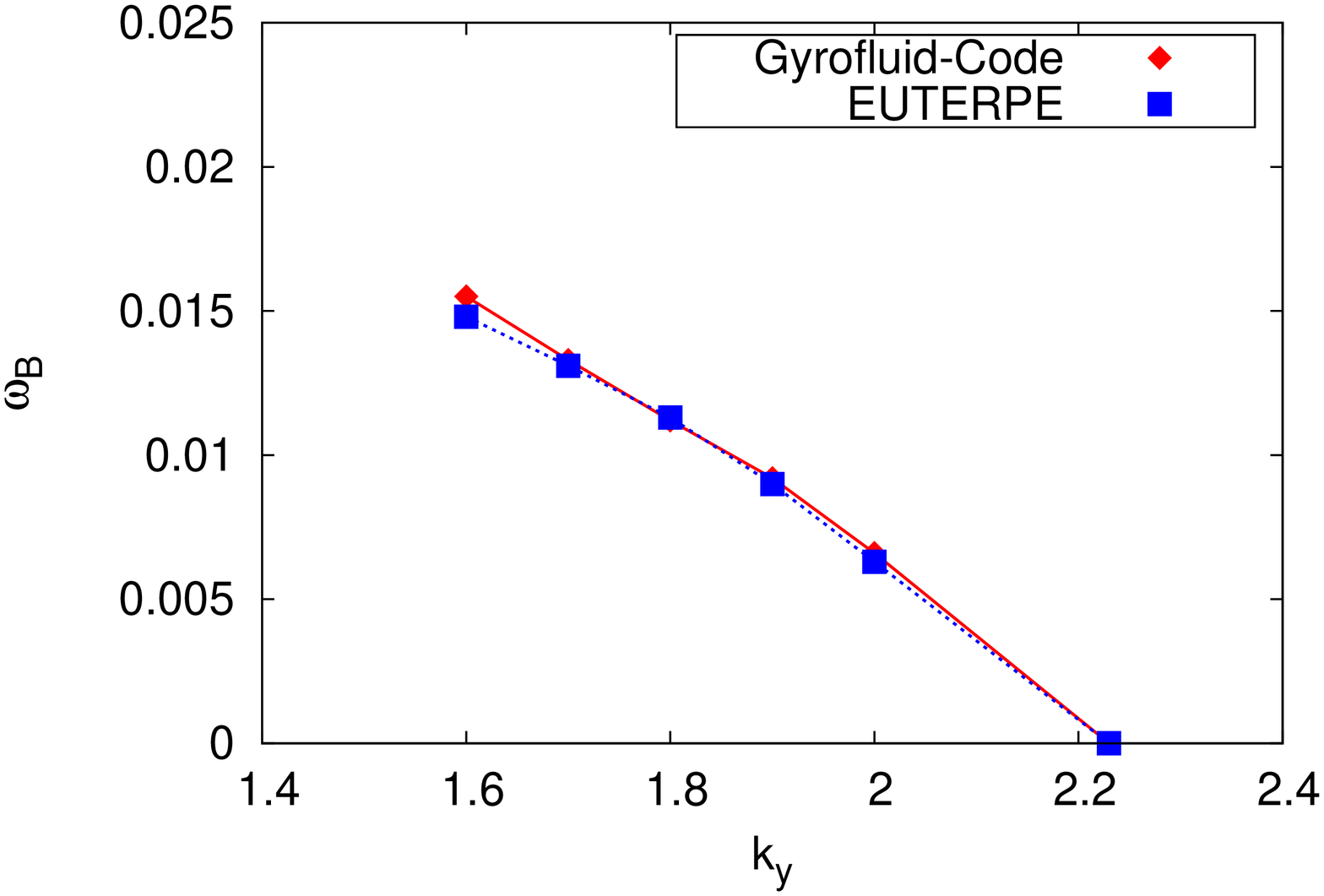}
\end{minipage}
\hspace{3cm}%\hfill
\begin{minipage}[hbt]{5cm}
	\centering
	\includegraphics[width=5cm,angle=-90]{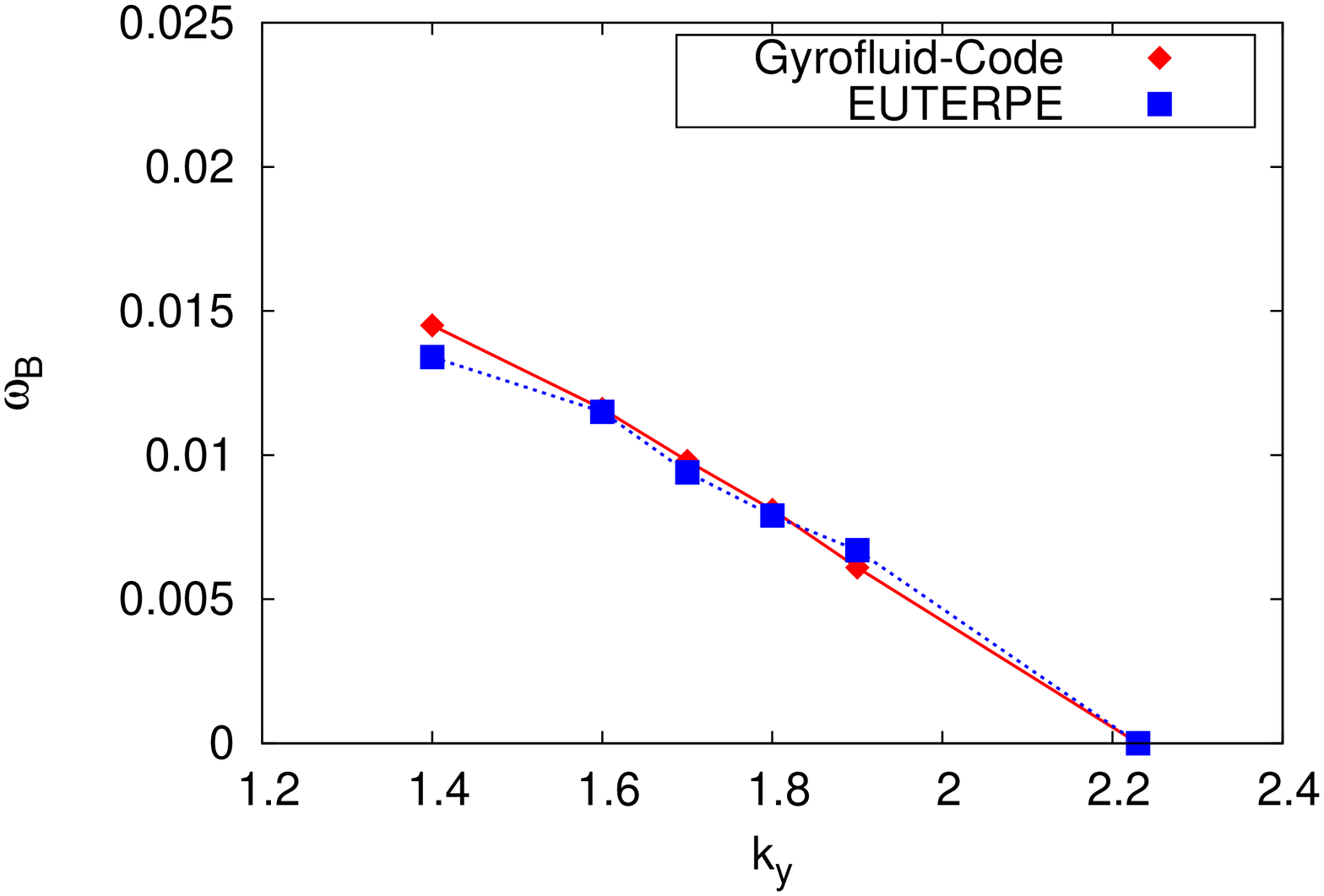}
\end{minipage}
\caption{Oscillation frequency as a function of the wavenumber $k_y$ for the two 
models using {the} parameter setup I (left panel) and
the parameter setup II (right panel).
 Both models deliver practically the same
 oscillation frequency in the saturated phase in the low- and medium-$\Delta^{\prime}$ range.}
\label{fig_bounce}
\end{figure}
In the gyrofluid simulations the oscillation frequency can always be clearly observed.
While for parameters of case I the frequency can be measured clearly with 
the gyrokinetic code EUTERPE, this is more difficult in case II. 
Therefore, to obtain good results the number
of markers was doubled to $N_p=3 \cdot 10^7$ and the previous time step 
was halfed to $\Delta\,t=0.125$. \\
The results are displayed in Figure~\ref{fig_bounce}.
The models agree very well for all wavenumbers $k_y$ shown here,
also for moderate values of $\Delta^{\prime} \sim 1$. These
results clearly show that also in this regime the 
oscillatory behaviour of the saturated reconnection process can be described 
completely by a fluid description. \\
From a rough kinetic estimation one gets $\omega_{\rm B}\sim k_y\,v_e\,w/\left(2\,l_s\right)$ 
~\cite{DL_1977_PRL, WCP_2005}, 
so the frequency is roughly proportional to the island width and 
the stability parameter $\Delta^{\prime}$ according to Eq.~(\ref{eq_drakenlin}). 
The results in Figure~\ref{fig_bounce}
confirm this linear scaling in the limit of low $\Delta^{\prime}$ values.

\subsection{Finite ion temperature effects} 
This section deals with the extension of previous nonlinear results by
including finite ion temperature effects using the full finite Larmor radius
(FLR) response.\\
We focus on the parameter case I and investigate the behaviour of the saturated island
half width and oscillation frequency 
with increasing ion temperature. \\
\begin{figure}[h]
\centering
\begin{minipage}[hbt]{5cm}
	\centering
	\includegraphics[width=5cm,angle=-90]{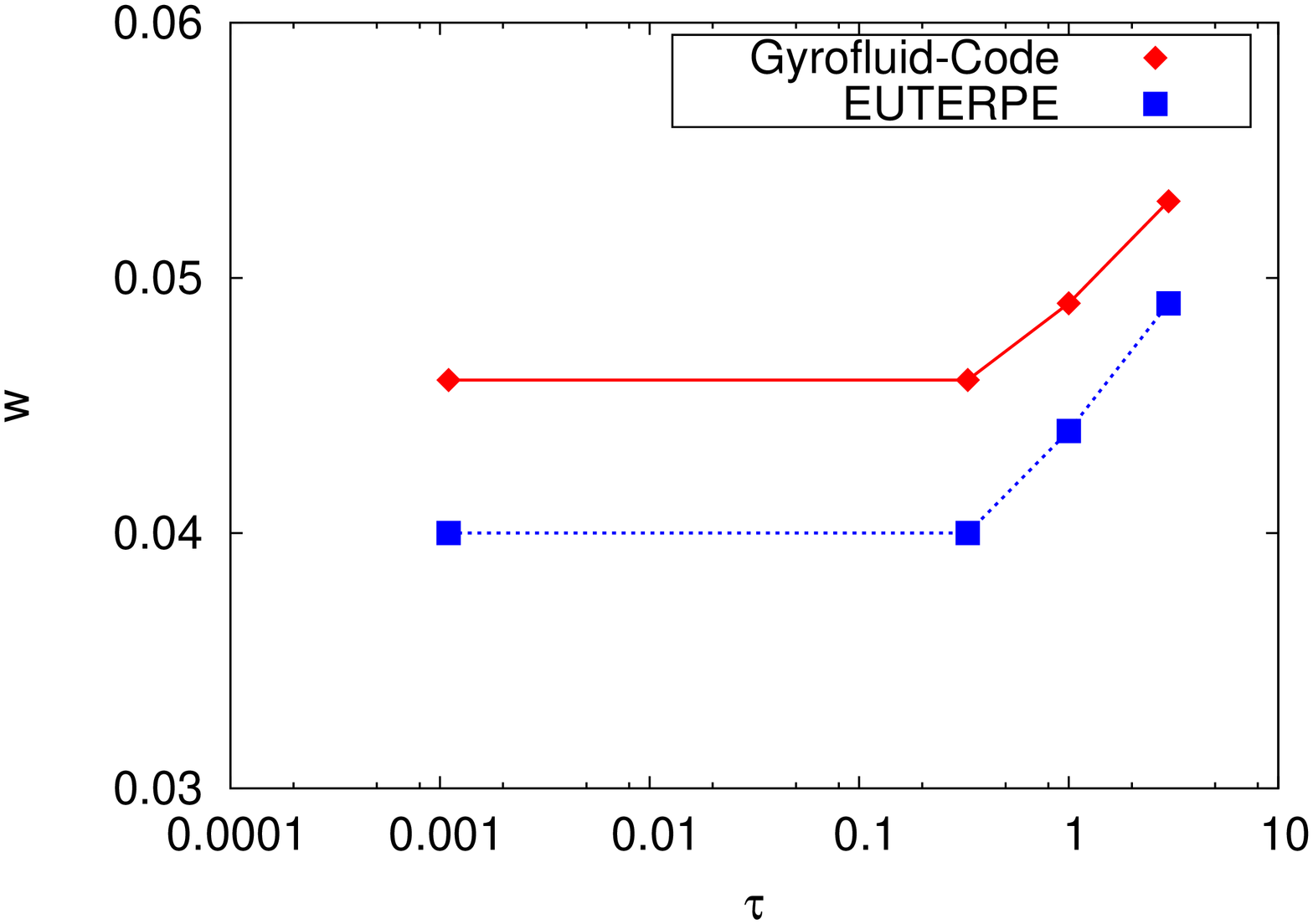}
	%\caption{Bild1}
	%\label{Bild1}
\end{minipage}
\hspace{3cm}%\hfill
\begin{minipage}[hbt]{5cm}
	\centering
	\includegraphics[width=5cm,angle=-90]{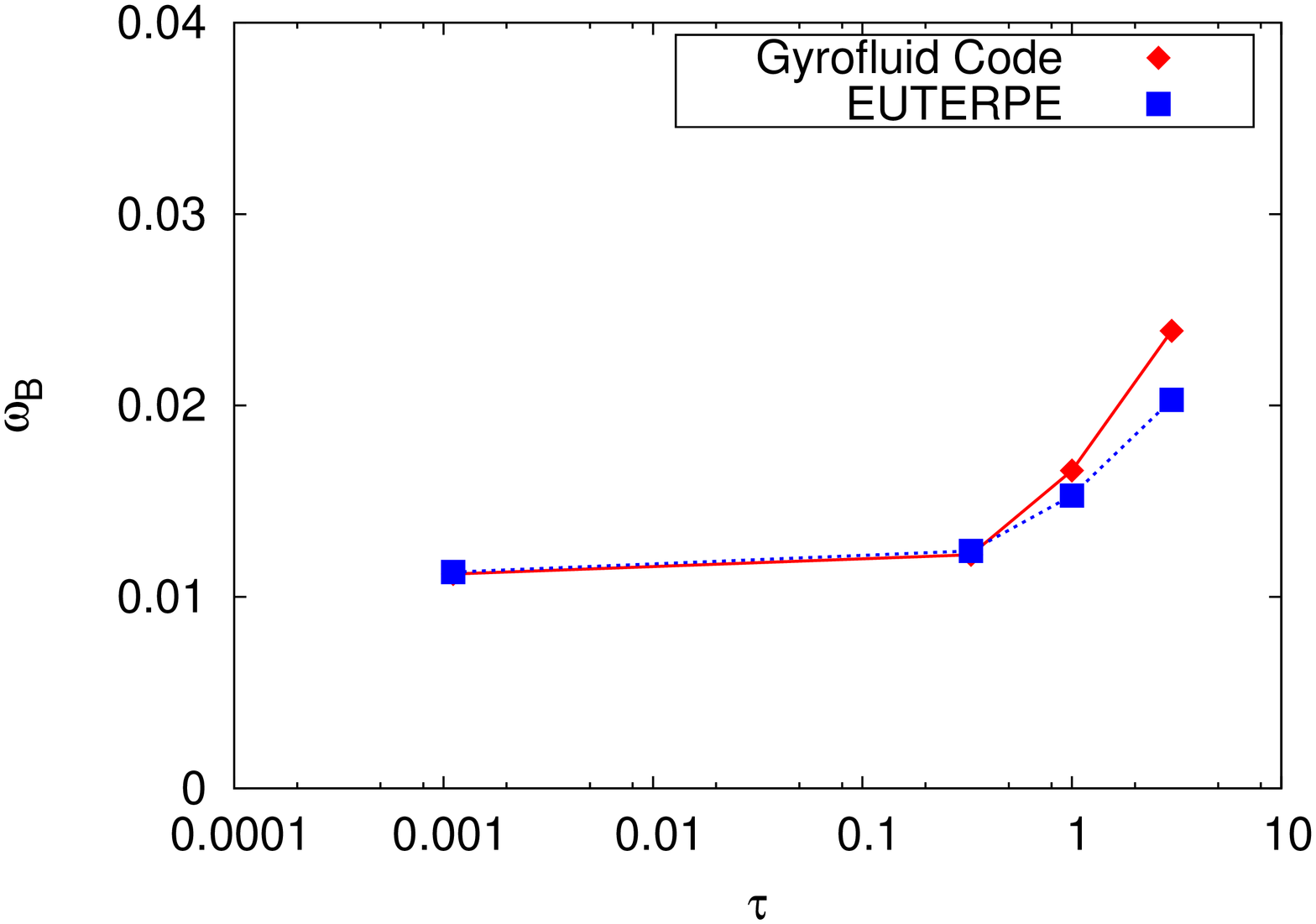}
	%\caption{Bild2}
	%\label{Bild2}
\end{minipage}
\caption{Comparison of the island half width $w$ (left panel) 
and the oscillation frequency $\omega_{\rm B}$ (right panel) as a function of the temperature ratio $\tau$
for the parameter Setup I and $k_y=1.8$.}
\label{fig_pade_caseI}
\end{figure}
In Figure~\ref{fig_pade_caseI}, left, the saturated island half width is shown when
the ion temperature is varied using the values $\tau={1/900,\,0.25,\,1,\,4}$ and fixing $k_y=1.8$.
The island width only changes by about
5\% over approximately three orders of magnitude of $\tau$.
This shows that finite Larmor radius effects on $w$ are weakly relevant 
for $\Delta^{\prime} \lesssim 1$. \\
As stated earlier, 
Ref.~\cite{DL_1977_PRL} predicts the general saturation condition $w \sim \delta$. 
Here, due to the influence of finite ion temperature, the parallel current channel width changes
 according to~\cite{Rogers_2007}
\begin{eqnarray}
\delta &\sim& \frac{\gamma\,l_s}{k_y\,v_e\,\sqrt{1+\tau}}.
\label{delta_rogers}
\end{eqnarray}
On the other hand the growth rate increases according to $\gamma \propto \sqrt{1+\tau}$,
as we have seen in section \ref{sec3}.
Using  Eq.~(\ref{eq_porc}) for the growth rate and 
Eq.~(\ref{delta_rogers}) for the modified current width, 
the generalized scaling of the saturated island half width 
for finite $\tau$ becomes 
\begin{eqnarray}
w&\sim&\Delta^{\prime}\,d_e^2
\end{eqnarray}
as stated for the drift kinetic case. This estimation makes evident that the saturated island 
width does not change significantly with ion temperature. \\
In contrast to the island half width, the oscillation frequency changes significantly 
when the temperature ratio is varied, as shown in the right panel of Figure~\ref{fig_pade_caseI}.
The dependence of the oscillation frequency
on the temperature ratio
 is similar to that of the growth rate.
However, even with small but finite $\tau \lesssim 1$, the two models agree completely
in the saturated phase.

\newpage

\section{Summary} \label{sec5}

We have simulated collisionless magnetic reconnection via the tearing instability 
with a gyrokinetic and a gyrofluid model. 
The results of both approaches have been compared to each other 
linearly and nonlinearly for an extended set of parameters. 
To the best of our knowledge, this is the first comparison of these two models 
for simulations of the collisionless tearing mode. \\
As a first step, we have applied a shooting method to benchmark the linear simulations 
of both codes in the drift kinetic limit.
The linear eigenmodes of the two models have been benchmarked for a single wave number 
and a fixed set of plasma parameters, whereas the linear growth rates of both codes 
have been compared for a range of wave numbers. It has been shown that in the linear regime 
both codes give results with high degree of accuracy.
Then the results of the two models have been compared over the whole spectrum 
of linearly unstable wave numbers for two sets of plasma parameters
showing a good agreement between the growth rates obtained with 
the gyrokinetic model and the gyrofluid one.\\
The linear simulations have been extended to the case of finite ion temperature, 
where we have shown that ion gyro-orbit averaging effects can be properly described 
by both approaches. 
Furthermore, numerical simulations in the small $\Delta' d_e$ range compare favorably 
 with the asymptotic theory by Porcelli {~\cite{P_1991}.} \\
Nonlinear simulations of both models have been carried out in the small-$\Delta'$ regime.
We have performed a detailed comparison of observables such as 
the evolution and saturation of the island width, as well as its oscillation frequency 
in the saturated phase, which has not been performed in this extend of parameter space so far. 
The gyrokinetic and gyrofluid simulations have shown that close to the marginal stability 
the evolution and saturation of the island width for both models is practically the same.
Moreover, an important and new observation is that the oscillation frequency 
of the island width shows no difference between the two models.
Therefore, the main result is that the nonlinear evolution of the collisionless tearing mode 
in the drift kinetic limit is essentially well described by the fluid theory.
We have also considered finite ion temperature effects in the saturated island phase. 
Here again both models differ only slightly when measuring the island width
and its oscillation frequency. Therefore, in the regimes investigated here, the
nonlinear reconnection physics can be completely described with a gyrofluid approach. \\
{Slightly} stronger deviations between the simulation results occur for $\Delta' \sim 1$, 
suggesting that further investigations will be of interest in this regime, 
as well as in cases where $\Delta' \gg 1$, for which a detailed nonlinear comparison
between the gyrokinetic and gyrofluid models is still missing. \\
%
\begin{comment}
The inclusion of kinetic effects in phase space that rely on 
resonant interactions of particles with the mode would not change significantly 
the collisionless tearing mode.
\end{comment}

\acknowledgments
{The authors would like to acknowledge fruitful discussions with Dario Borgogno and Alessandro Zocco.
This work was supported by the European Community under the contracts of Association between Euratom and ENEA 
and by the Euratom research and training programme 2014--2018. 
Part of this work was carried out using the HELIOS supercomputer system at the 
Computational Simulation Centre of International Fusion Energy Research Centre (IFERC-CSC), Aomori, Japan, 
under the Broader Approach collaboration between Euratom and Japan, implemented by Fusion for Energy and JAEA.}
The views and opinions expressed herein do not necessarily reflect those of the European Commission. \\

\end{document}